\pdfoutput=1
\documentclass{aa}
\usepackage{graphicx}
\usepackage{txfonts}
\usepackage[colorlinks=true,citecolor=blue]{hyperref}
\usepackage{natbib}
\bibpunct{(}{)}{;}{a}{}{,}
\usepackage{color}
\usepackage{subfig}

\newcommand{\kms}{\mbox{km\,s$^{-1}$}}
\newcommand{\Macc}{\mbox{$\dot{M}_\textup{acc}$}}
\newcommand{\Teff}{\mbox{$T_\textup{eff}$}}

\newcommand{\Lsun}{\mbox{$L_\sun$}}
\newcommand{\Rsun}{\mbox{$R_\sun$}}
\newcommand{\Msun}{\mbox{$M_\sun$}}
\newcommand{\Lstar}{\mbox{$L_*$}}
\newcommand{\Rstar}{\mbox{$R_*$}}
\newcommand{\Mstar}{\mbox{$M_*$}}
\newcommand{\vsini}{\mbox{$v \sin i$}}

\begin{document} 


   \title{Accretion in low-mass members of the Orion Nebula Cluster with young transition disks}


   \author{R. M. G. de Albuquerque\inst{1,2,3}
          \and
          J. F. Gameiro\inst{1,2}
          \and
          S. H. P. Alencar\inst{4}
          \and
          J. J. G. Lima\inst{1,2}
          \and
          C. Sauty\inst{3,5}
          \and
          C. Melo\inst{6}
          }

   \institute{Instituto de Astrof\'isica e Ci\^encias do Espa\c{c}o, Universidade do Porto, CAUP, Rua das Estrelas, PT4150-762 Porto, Portugal \\ \email{Raquel.Albuquerque@astro.up.pt}
              \and
              Departamento de F\'isica e Astronomia, Faculdade de Ci\^encias, Universidade do Porto, Rua do Campo Alegre 687, PT4169-007 Porto, Portugal
              \and         
              Laboratoire Univers et Th\'eories, Observatoire de Paris, Universit\'e PSL, CNRS, Universit\'e de Paris, 92190 Meudon, France
              \and         
              Departamento de F\'isica - ICEx-UFMG, Av. Ant\^onio Carlos, 6627, 30270–901 Belo Horizonte, MG, Brazil
              \and
              LUPM, Universit\'e de Montpellier, UMR 5299 CNRS/IN2P3, cc072, place Eug\`ene Bataillon, 34090 Montpellier, France
              \and              
              European Southern Observatory, Alonso de C\'ordova 3107, Vitacura, Regi\'on Metropolitana, Chile
             }
             
   \date{Received ...; accepted ...}


 
  \abstract
   {Although the Orion Nebula Cluster is one of the most studied clusters in the solar neighborhood, the evolution of the very low-mass members (\Mstar < 0.25 \Msun) has not been fully addressed due to their faintness.}
   {Our goal is to verify if some young and very low-mass objects in the Orion Nebula Cluster show evidence of ongoing accretion using broadband VLT/X-Shooter spectra.}
   {For each target, we determined the corresponding stellar parameters, veiling, observed Balmer jump, and accretion rates. Additionally, we searched for the existence of circumstellar disks through available on-line photometry.}
   {We detected accretion activity in three young stellar objects in the Orion Nebula Cluster, two of them being in the very low-mass range. We also detected the presence of young transition disks with ages between 1 and 3.5 Myr.}
   {}


   \maketitle

\section{Introduction}

The Orion Nebula Cluster (ONC) is one of the nearest and massive star forming clusters that belongs to the Orion A molecular cloud, a region known for its young stellar population \citep{kounkel2017}. Estimates indicate that this cluster has an age of approximately 2.2 Myr with a scatter of few Myr \citep{reggiani2011}, in agreement with previous studies \citep{hillenbrand1997,dario2010}. The Orion Nebula (M42 or NGC 1976), extends up to a radius of approximately 3 pc. The densest region in the $\sim 0.3$ pc core of the ONC corresponds to the Trapezium cluster \citep{hillenbrand1997,hillenbrand2013}, where massive young stars excite and illuminate the nebula. Some of the young stellar objects (YSOs) in the ONC are visible in the optical due to the strong radiation field and the \ion{H}{II} region inflation created by a newborn O-type star, $\theta^1$ C Ori, which cleaned part of the cluster \citep{muench2008,pettersson2014}. It is a region of interest for research in star formation, in particular to analyze ongoing accretion processes in pre-main sequence (PMS) stars.

Important contributions to the study of the ONC have been made, especially with large surveys such as \cite{hillenbrand1997} and \cite{dario2010,dario2012,dario2016}, for instance. Through photometry and spectroscopy techniques, they characterized more than one thousand ONC members. But one of the missing aspects here is the characterization of the very low-mass YSOs in terms of accretion activity. Accretion has a key role in determining not only the mass of these objects, but also their angular momentum.

In order to explain the accretion mechanisms observed in these active YSOs, magnetospheric accretion models have been adopted. These models suggest the presence of strong magnetic fields, which truncate the inner region of the disk and enable the transport of material from the disk to the star through accretion columns. The regions where the material shocks onto the stellar surface are the so-called hot spots, where ultra-violet energy is released. This energy is fingerprinted in the star spectrum as a continuum excess emission known as veiling.
These magnetic fields are also responsible for carrying angular momentum from the star to the disk leading to the deceleration of the YSO \citep{bouvier2007}. 

Although there is some progress concerning the study of the characterization of the magnetic fields in low-mass stars (i.e., \cite{donati2010,johnstone2013,hill2019}), very little is known about the very low-mass YSOs, due to their faintness. Hopefully, new instruments such as SPIRou \citep{thibault2012} will have the potential to disentangle the magnetic field topologies for YSOs in the very low-mass range.

In a previous work of \cite{biazzo2009}, about 90 very low-mass members of the ONC were analyzed with the multi-fiber spectrograph FLAMES. Using the \cite{white2003} accretion criterion, none of those objects were accreting. The criterion states that an object is accreting if the H$\alpha$ width at 10\% of peak intensity is greater than 270 \kms. Nevertheless, less than 10\% of those objects showed a 10\% width larger than the median of the sample. Additionally, those objects also showed slow rotation rates, meaning that the star is still locked to the disk, Li I (670.8 nm) in absorption, indicative of their youth, and an infrared excess suggesting the presence of a circumstellar disk. 

The aim of this study is to probe for accretion among the young and very-low mass ONC members. The opportunity to observe these PMS stars with the X-Shooter instrument, installed at the Very Large Telescope (VLT), allowed us to analyze accretion tracers in three different wavebands simultaneously. Additionally, it gave us more information about accretion processes in very-low mass YSOs.

The paper is structured as follows. In Sect. \ref{sect:sample}, we present the sample, characterize the observations, and describe how the reduction was performed. In Sect. \ref{sect:parameters}, we show the results of the derived stellar parameters for each target; and in Sect. \ref{sect:analysys}, we analyze if the targets have ongoing accretion activity. In Sect. \ref{sect:discussion}, we discuss the obtained results and compare them with previous works, if available. In Sect. \ref{sect:conclusions}, we address the main conclusions of this study.

\section{Sample, observations and data reduction}\label{sect:sample}

\begin{figure}
   \centering
   \includegraphics[width=0.8\hsize]{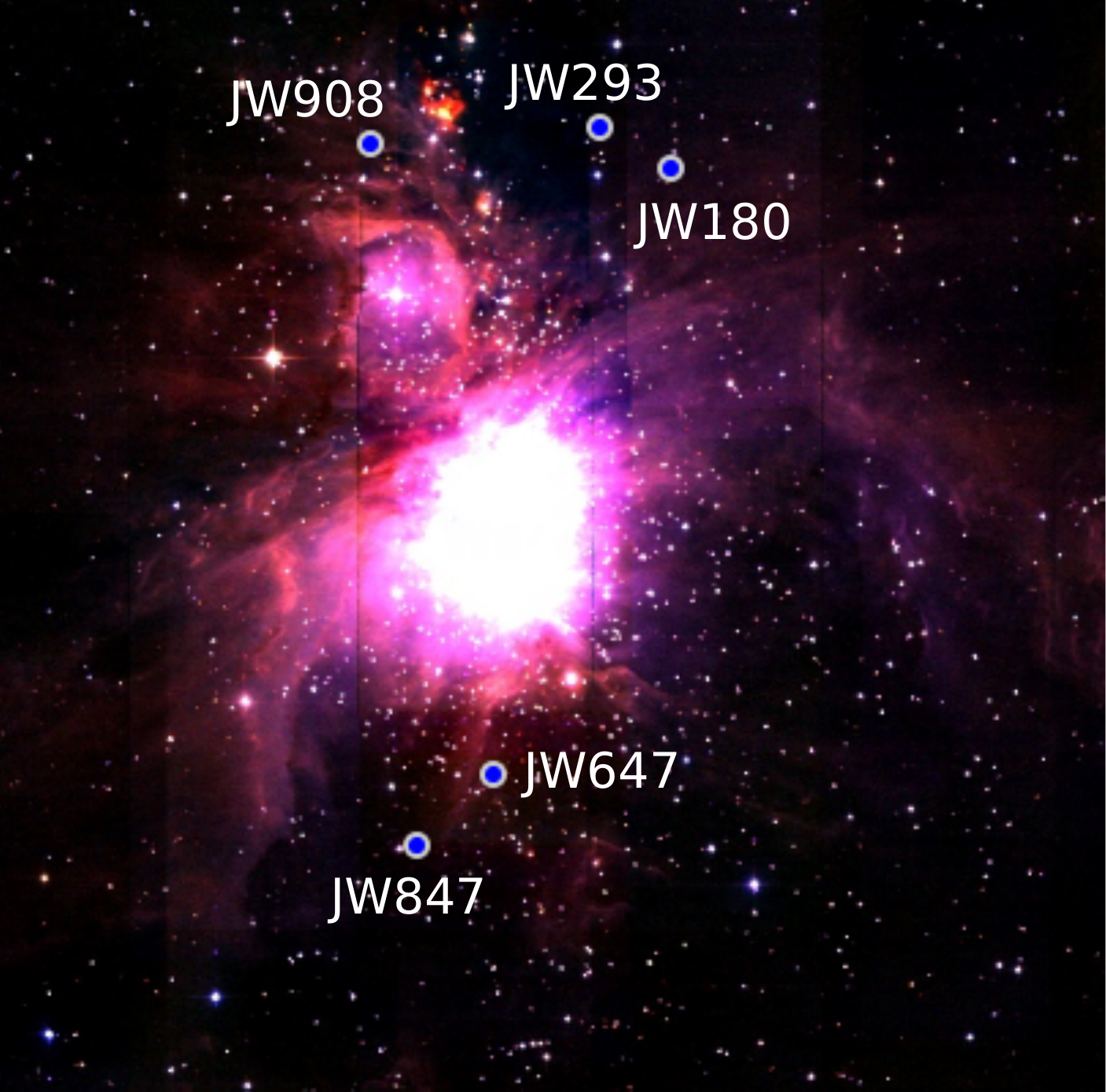}
      \caption{Distribution of the targets in the ONC. Figure made with Aladin Lite interactive sky maps \citep{bonnarel2000,boch2014}.}
         \label{fig:ONC_targets03}
   \end{figure}

The sample is composed of five  objects belonging to the ONC (see Fig.\ref{fig:ONC_targets03}) listed in Table \ref{table:1}.
Based on a previous target selection by \cite{biazzo2009}, including very low-mass ONC members, it was not clear if some of those objects were accreting. For such matter, the selected targets are some of the brightest stars in the $I_C$ band according to \cite{biazzo2009}. 
The star JW847 was not in the target list of this last study, but was later included in the observational proposal. These five targets were chosen in order to confirm if they have ongoing accretion activity and how their study can improve our understanding of disk-locking mechanisms in the very low-mass range.
These objects have a membership probability equal to or higher than $97 \%$, determined by \cite{jones1988}, who measured the proper motions for over a thousand objects in the Orion Nebula.
In the same table, we also list the 2MASS name, the distance, spectral type, radial velocity, and projected rotational velocity available in the literature for each target.

The distances were computed from the parallaxes available in Gaia DR2 \citep{gaia2016,gaia2018,luri2018}. In general, these values are smaller than the average distance estimated to the ONC (e.g., $414 \pm 7$ pc by \cite{menten2007}), especially for JW180. In order to examine the validity of the parallax measurements, we checked specific Gaia DR2 source parameters that suggest that the parallax for JW180 is the least reliable among the stars in the sample. However, the stellar parameters computed for JW180 do not change significantly if we consider an average distance to the ONC. More details concerning the values of the Gaia DR2 source parameters and the impact of the distance in the derived stellar parameters of JW180 can be found in Appendix \ref{app:distances}.

\begin{table*}
\caption{Stellar parameters available in the literature for the ONC targets. The targets are listed in the first column, followed by the corresponding 2MASS ID, the distance to the target ($d$), spectral type (SpT), radial velocity (RV) and rotational velocity (\vsini).}   
\label{table:1}     
\centering                         
\begin{tabular}{c c c c c c}       
\hline\hline                 
JW & 2MASS & $d^{(1)}$ (pc) & SpT$^{(2)}$ & RV$^{(3)}$ (\kms) & \vsini$^{(4)}$ (\kms) \\  
\hline                        
   180 & J05345819-0511536 & 323 $\pm$ 35 & M5    & 20.0 $\pm$ 1.4 & 44.1 $\pm$ 1.0 \\      
   293 & J05350682-0510385 & 386 $\pm$ 12 & M5    & 27.9 $\pm$ 1.7 & 46.8 $\pm$ 3.0 \\
   647 & J05352029-0530395 & 412 $\pm$ 7 & M5e   & 26.4 $\pm$ 0.3 & 15.0 $\pm$ 0.7 \\
   847 & J05352983-0532534 & 386 $\pm$ 8 & K3/G8 & 27.8 $\pm$ 0.2 & 44.5 $\pm$ 0.7 \\
   908 & J05353534-0511114 & 395 $\pm$ 14 & M4.5  & 29.4 $\pm$ 0.6 & 16.3 $\pm$ 0.6 \\ 
\hline                                  
\end{tabular}
\tablebib{(1)~\citet{gaia2016,gaia2018,luri2018}; (2)~\citet{hillenbrand1997};(3)~\citet{cottaar2015};(4)~\citet{dario2016}.}
\end{table*}

The observations were taken in service mode at the ESO/VLT X-shooter spectrograph \citep{vernet2011} between January and March of 2015\footnote{The corresponding data can be found at the ESO archive with the program ID 094.C-0327(A) (PI: S. Alencar).}.

The X-shooter instrument covers the spectral range from 300 to 2500 nm, approximately, including three arms in different wavebands: ultra-violet (UVB), visible (VIS) and near-infrared (NIR). The UVB arm extends from 300 to 550 nm, while the VIS arm ranges between 550 to 1050 nm, and the NIR arm goes from 1050 to 2500 nm \citep{vernet2011}. 

The targets were observed in slit-stare mode with slit widths of 1.3\arcsec, 1.2\arcsec, and 1.2\arcsec, corresponding to a resolution of 4000, 6700, and 3900 for the UVB, VIS, and NIR arms, respectively. The observations were done in the stare mode instead of the requested nodding mode, which made the reduction process a complex task due to the difficulty of subtracting the sky in the NIR arm.

The spectra of all arms were reduced with version 2.7.1 of X-shooter pipeline \citep{modigliani2010}. The 2D outputs were later extracted with the IRAF\footnote{IRAF is distributed by the National Optical Astronomy Observatory, which is operated by the Association of Universities for Research in Astronomy (AURA) under a cooperative agreement with the National Science Foundation.} task apall. The resulting 1D spectra were corrected from telluric absorptions using the standard tellurics observed, before or after each target, with IRAF task telluric.
An additional correction for the flux in the NIR was made in order to overlap the end of the VIS arm with the beginning of the NIR arm.

\section{Stellar parameter determination}\label{sect:parameters}

The determination of stellar parameters in YSOs is not straightforward. Some issues that may influence the spectral type assignment include nebular emission or inadequate background subtraction, including both effects of the Earth sky and background from the ONC \citep{sicilia-aguilar2005,dario2009,hillenbrand2013}.

Reddening effects should also be considered, namely extinction due to interstellar dust and the existence of circumstellar disks \citep{manara2013b}. Additionally, if the targets have a reasonable accretion activity, the continuum excess will be enough to make the spectrum bluer and lead us to think that we are dealing with an earlier type star. The reverse situation may occur due to the presence of cool spots in non-accreting YSOs that can induce late-type features in the spectra \citep{hillenbrand2013}.
In order to achieve a correct spectral type determination and corresponding effective temperature, all of these effects should be taken into account.

\subsection{Extinction and spectral type}

Before assigning spectral types, we should account for the visual extinction ($A_V$) associated with each star. Nebulae, like the ONC, are rich in gas and dust. Between the star and the observer, these components can absorb and scatter the light that is emitted by the target. In order to correct the spectra from these effects, we determine the extinction for each target.

In order to compute $A_V$, we used the spectra of the VIS arm and compared it with templates selected from the library  of non-accreting class III YSOs from \cite{manara2013b} and \cite{manara2017}. This selection was made according to the original classification of the targets listed in Table \ref{table:1}. The nearby spectral types are used as our first guesses.

Firstly, both targets and templates were normalized with the flux measured at 750 nm. 
Secondly, we reddened the templates from 600 to 800 nm, with extinctions ranging from 0 to 5 mag in steps of 0.1. 
Although the interstellar medium is not uniform, we assumed the same extinction law for all the targets and test two values for the reddening parameter $R_V$.
For that, we used the extinction law from \cite{cardelli1989}, assuming a reddening parameter of $R_V=3.1$ and 5.5, two values commonly used for ONC members \citep{dario2012,dario2016,manara2012,hillenbrand2013}. 

\begin{figure*}
   \centering
   \includegraphics[width=1.0\hsize]{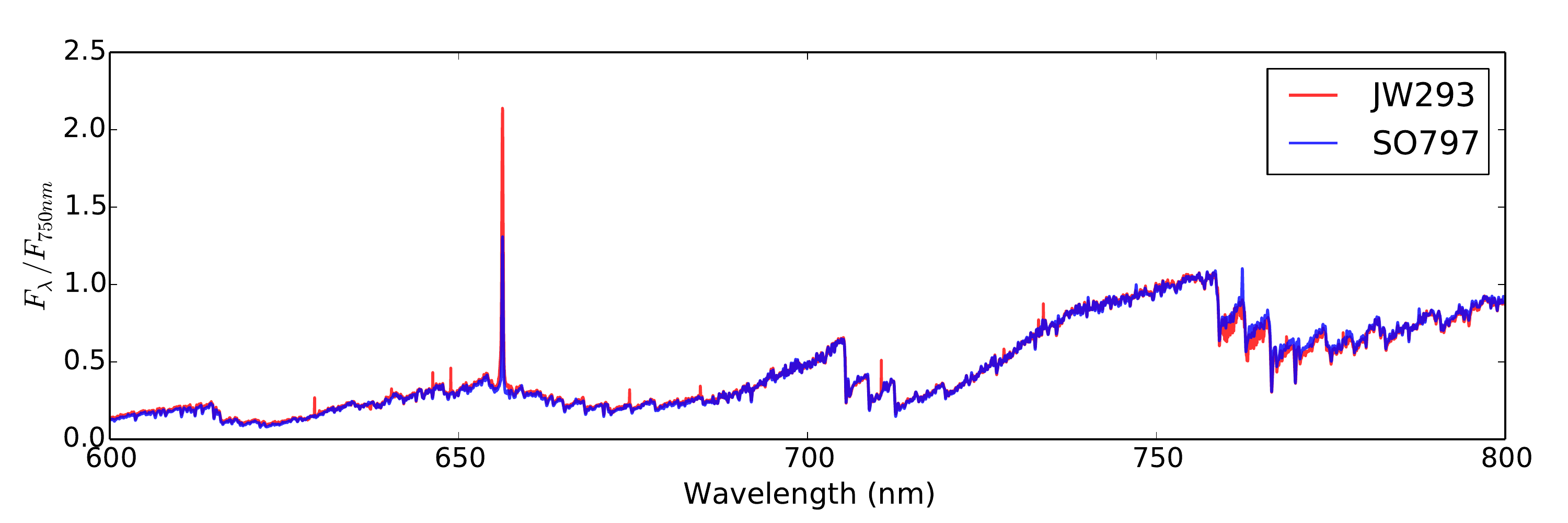}
      \caption{Comparison between spectrum of JW293, corrected from visual extinction, and corresponding template SO797 assuming $R_V=3.1$.}
         \label{fig:Av_JW293_SO797}
   \end{figure*}

Next, we compared the reddened template with the target. The comparison is performed with \texttt{lmfit} \citep{newville2014}, built on Levenberg-Marquardt algorithm. The best fit with the lowest $\chi^2$ value returns the extinction value corresponding to the target. All the targets show $A_V < 0.1$ mag for both $R_V$ values considered. Like in \cite{manara2017}, we can consider that $A_V < 0.5$ mag are negligible, or very low, extinction values.
One of the results can be seen in Fig. \ref{fig:Av_JW293_SO797}, where the spectrum of JW293 is shown after being extinction corrected (red line) and is compared with the corresponding template (blue line).

Determining the spectral type is the starting point to derive not only the effective temperature, but also the remaining stellar parameters that characterize the star. The visible waveband is quite important for this task, since it includes relevant molecular and atomic absorption features that are sensitive to temperature \citep{hillenbrand2013}.

The spectral type of the five targets was previously determined in \cite{hillenbrand1997}, most of them being mid M-type stars (see Table \ref{table:1}). Two of them were reclassified in \cite{hillenbrand2013}: JW180 was confirmed to be an M5, but JW647 was changed from an M5 to an M0.5.
The spectra of M-type stars are characterized by the presence of metal oxide species. These species include bands of titanium oxide (TiO) and vanadium oxide (VO), whose strengths increase with decreasing temperature. For the case of K-type stars, these molecular bands will be less intense towards earlier spectral types \citep{gray2009,riddick2007b}. 

\begin{table*}
\caption{Spectral indices used in this work. All spectral indices are calculated from the ratio between the average flux of the numerator (N) and denominator (D), except the following cases: \tablefoottext{a}{$\frac{F(N)}{F(D)}\frac{F_{line}(4650)}{F_{line}(5100)}$;}
\tablefoottext{b}{$\log\left ( \frac{F(N)}{F(D)}-1  \right )$}.}
\label{table:indices}     
\centering                      
\begin{tabular}{c c c c c}       
\hline\hline                
Index & Range of validity & Numerator (\AA) & Denominator (\AA) & Ref. \\ 
\hline                        
   R1       & M2.5-M8 & 8025-8130 & 8015-8025 & 1 \\      
   R2       & M3-M8   & 8145-8460 & 8460-8470 & 1 \\
   R3       & M2.5-M8 & (8025-8130)+(8415-8460) & (8015-8025)+(8460-8470) & 1 \\
   TiO 8465 & M3-M8   & 8405-8425 & 8455-8475 & 1 \\
   VO 2     & M3-M8   & 7920-7960 & 8130-8150 & 1 \\ 
   VO 7445  & M5-M8   & 0.5625(7350-7400)+0.4375(7510-7560) & 7420-7470 & 1 \\
   R5150\tablefootmark{a} & K0-M0 & 5050-5150 & 4600-4700 & 2 \\
   TiO 7140 \tablefootmark{b} & M0-M4.5 & 7005-7035 & 7130-7155 & 2 \\
   TiO 7700 & M3-M8 & 8120-8160 & 7750-7800 & 2 \\
   TiO 8465 & M4-M8 & 8345-8385 & 8455-8475 & 2 \\
   TiO 7140 & K5-M6 & 7020-7050 & 7125-7155 & 3 \\ 
\hline                                  
\end{tabular}
\tablebib{(1)~\citet{riddick2007b}; (2)~\citet{herczeg2014}; (3)~\citet{jeffries2007,oliveira2003}.}
\end{table*}

In order to assign spectral types, we calculated narrow-band spectral indices for each target, by computing ratios of average fluxes in the optical waveband of the spectrum for the targets. The indices for each spectral type will depend on the strength of specific molecular lines that are temperature-sensitive. For the late M-type stars, we used \cite{riddick2007b} and spectral indices, which are independent of reddening and nebular emission. While for K and early M-type stars, we used spectral indices from \cite{herczeg2014}. Additionally, we computed the TiO (7140 \AA) spectral index by \cite{jeffries2007} that is valid for stars with spectral types K5 and later. All the selected indices are listed in Table \ref{table:indices}. 

The selected spectral type for each star is obtained after averaging the spectral types calculated from the previous indices. We confirm the spectral type by comparing the target spectra with templates of similar spectral type taken from the library of non-accreting class III YSOs \citep{manara2013b,manara2017}.
In order to accomplish that, we normalized VIS spectra at 750 nm and applied a gaussian instrumental broadening\footnote{We used the function \texttt{broadGaussFast} available at \url{https://github.com/sczesla/PyAstronomy}.} to the templates in order to get the targets resolution. A list with the best template matches can be found in Table \ref{table:templates}.

We concluded from the comparison that JW180 and JW908 are reasonably well fit by the corresponding templates. For JW293, when comparing with M4.5 and M5 templates, we found a better fit for the M4.5 template. The template match to the earliest types, JW647 and JW847, was not satisfactory.
Although we retrieved for JW647 an average spectral type of approximately M2.0, an M1 template seems to better fit the target spectrum. For JW847, the R5150 index points to  K1.2, but the TiO index indicates that this star should be a K6.6. Interestingly, through \cite{tripicchio1997} empirical relation with the \ion{Na}{I} doublet, the results indicate that the spectral type of JW847 should be between a K0 and a K1. However, from the comparison with the library spectra, the best match is obtained with a K6 template. We found that the K0 and K1 templates are too hot to fit the target, and a K7 template presents more  pronounced molecular features. In view of this, we adopted, for both stars, the spectral classes obtained from templates comparison.

In Table \ref{table:stellarparams}, we list the spectral types derived with the spectral indices from different works: \cite{riddick2007b} (RRL07), \cite{jeffries2007} (J07) and \cite{herczeg2014} (HH14) and the adopted spectral type which corresponds to the best matched template. To convert from spectral type to temperature, we used the temperature scale obtained by \cite{kenyon1995} and \cite{luhman2003} for earlier- and M-type stars, respectively. The corresponding effective temperatures for each star are shown as well.
For the spectral type, we assumed an error of one subclass, which corresponds to approximately 100 K.

\begin{table}
\caption{List of templates used in this work. }             
\label{table:templates}     
\centering                         
\begin{tabular}{l c c c}       
\hline\hline                 
Template & 2MASS & $d$ (pc)$^{(1)}$ & SpT$^{(2)}$  \\  
\hline                        
   Par-Lup3-2      & J16083578-3903479 & 166 $\pm$ 3 & M5   \\      
   SO797           & J05385492-0228583 & 354 $\pm$ 16 & M4.5 \\
   TWA13B          & J11211745-3446497 & 60  $\pm$ 0.1& M1  \\
   RX J1543.1-3920 & J15430624-3920194 & 166 $\pm$ 7 & K6$^{(3)}$ \\
\hline                                  
\end{tabular}
\tablebib{(1)~\citet{gaia2016,gaia2018,luri2018}; (2)~\citet{manara2013b}; (3)~\citet{manara2017}.}
\end{table}

\begin{table*}
\caption{Stellar parameters derived for the ONC targets. The spectral types derived in this work through spectral indices of \citet{riddick2007b} (RRL07), \citet{jeffries2007} (J07), and \citet{herczeg2014} (HH14) are listed, as well as the adopted one. The stellar parameters derived include effective temperature (\Teff), stellar luminosity (\Lstar), stellar radius (\Rstar), stellar mass (\Mstar), age, veiling measured at 610 nm ($r_{610\textup{nm}}$), and the observed Balmer jump ($BJ_\textup{obs}$). For the \Teff~an error of 100K is considered, and for the veiling measurements, the error is less than 0.2.}            
\label{table:stellarparams}     
\centering                         
\begin{tabular}{c c c c c c c c c c c c}       
\hline\hline                 
JW & \multicolumn{4}{c}{SpT derived in this work} & \Teff (K) & \Lstar (\Lsun) & \Rstar (\Rsun) & \Mstar (\Msun) & Age (Myr) & $r_{610\textup{nm}}$ & $BJ_\textup{obs}$\\  
   & RRL07$^{(1)}$ & J07$^{(2)}$ & HH14$^{(3)}$ & Adopted & & & & & & \\  
\hline                        
   180 & M5.0 & M5.1 & M5.0 & M5 & 3125 & 0.11 $\pm$ 0.05 & 1.12 $\pm$ 0.24 & 0.17 $\pm$ 0.04 & 3.53 $\pm$ 2.15 & 0.0 & 0.5 \\
   293 & M4.8 & M4.8 & M4.8 & M4.5 & 3197 & 0.13 $\pm$ 0.05 & 1.17 $\pm$ 0.21 & 0.21 $\pm$ 0.04 & 3.14 $\pm$ 1.02 & 0.0 & 0.5 \\
   647 & M3.2 & M0.6 & M2.1 & M1 & 3705 & 0.30 $\pm$ 0.12 & 1.32 $\pm$ 0.21 & 0.47 $\pm$ 0.07 & 3.26 $\pm$ 2.17 & 0.5 & 1.1 \\
   847 & ...  & K6.6 & K1.2 & K6 & 4205 & 1.93 $\pm$ 0.78 & 2.62 $\pm$ 0.21 & 0.90 $\pm$ 0.13 & 1.03 $\pm$ 0.67 & 0.0 & 0.6 \\
   908 & M4.5 & M4.5 & M4.4 & M4.5 & 3197 & 0.14 $\pm$ 0.06 & 1.21 $\pm$ 0.21 & 0.21 $\pm$ 0.04 & 2.99 $\pm$ 1.10 & 0.0 & 0.5 \\
\hline                                  
\end{tabular}
\tablebib{(1)~\citet{riddick2007b}; (2)~\citet{jeffries2007}; (3)~\citet{herczeg2014}.}
\end{table*}

\subsection{Stellar luminosity and radius}

The stellar flux for each target was determined following the procedure in \cite{manara2013b}. Firstly, we integrated the flux of the entire X-shooter spectrum from 350 to 2450 nm, excluding the  first and last 50 nm of noisy data and substituting the two strongest $H_{2}O$ telluric regions in the NIR (at 1330-1550 nm and 1780-2080 nm) via linearly interpolated straight lines. Secondly, BT-Settl synthetic spectra from \cite{allard2012} were used to extend the target spectrum, shortward of 350 nm and longward of 2450 nm. The synthetic patches are scaled to the flux values at these two last wavelengths. Then, the "synthetic flux" is integrated and added to the observed one. The synthetic spectra chosen have a compatible effective temperature with the targets, solar metallicity \citep{dorazi2009}, and $\log g = 4$, a common value adopted for low-mass YSOs. 

A flux correction is performed whenever the continuum veiling is not negligible (see Sect. \ref{sect:veiling}) to subtract the non-photospheric excess continuum flux from the total flux.
The stellar flux ($F_*$) is then converted in luminosity ($L_*$) through $L_* = 4 \pi d^2 F_*$, where $d$ is the distance to the star. In this work, we derived the distances from the Gaia DR2 parallaxes \citep{gaia2016,gaia2018,luri2018}, listed in Table \ref{table:1}, and we estimated the stellar radius ($R_*$) through the equation
\begin{equation}
    R_* = \left (  \frac{F_* d^2}{\sigma \Teff^4} \right )^{1/2} \, ,
\end{equation}
where $\sigma$ is the Stephan-Boltzmann constant. The luminosities and stellar radii derived  are shown in Table \ref{table:stellarparams}.

\subsection{Stellar mass and age}
\begin{figure}
   \centering
   \includegraphics[width=\hsize]{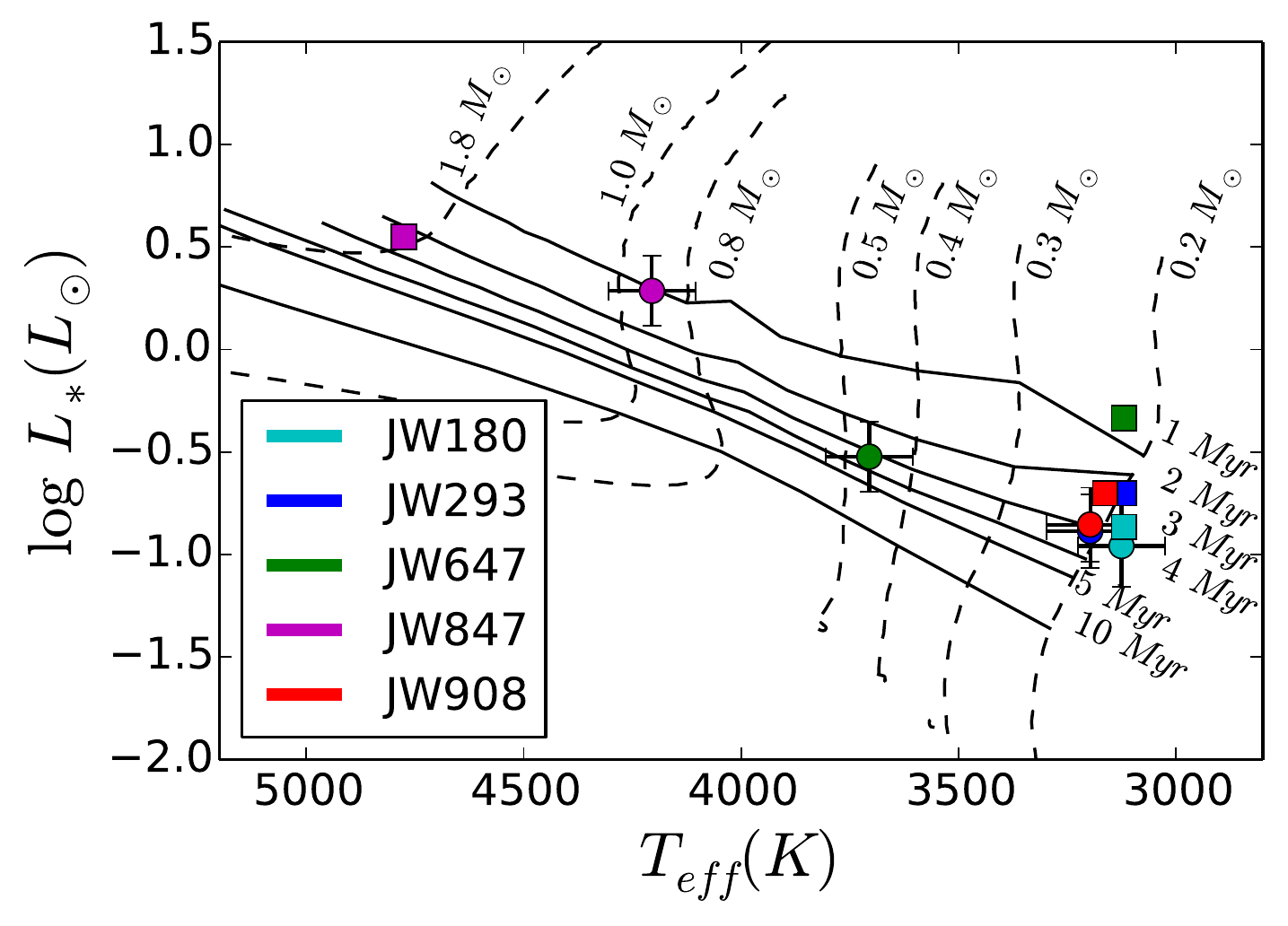}
      \caption{Hertzsprung-Russel diagram with \cite{siess2000} evolutionary tracks from 0.2 to 1.8 $M_\sun$ (dashed lines). Isochrones corresponding to 1, 2, 3, 4, 5, and 10 Myrs (solid lines) are also represented. The circles correspond to the values determined in this work and the squares to the values of \cite{hillenbrand1997}, rescaled to the distances derived from the Gaia DR2 parallaxes.}
         \label{HRD_Siess2000}
   \end{figure}

In Fig. \ref{HRD_Siess2000}, we plot the position of our targets in the Hertzsprung-Russel (HR) diagram. We also include the \cite{siess2000} pre-main sequence evolutionary tracks for several masses and the isochrones from 1 to 10 Myrs considering a solar metallicity ($Z = 0.020$, $X = 0.703$ and $Y = 0.277$).
The circles correspond to the effective temperatures and stellar luminosities determined in this work and listed in Table \ref{table:stellarparams}.
The squares correspond to the values of \cite{hillenbrand1997}, whose fluxes were rescaled to distances derived from the Gaia DR2 parallaxes \citep{luri2018}.

In Table \ref{table:stellarparams}, we also list the stellar masses and ages derived by the \cite{siess2000} online tool. By giving the effective temperature and stellar luminosity as inputs, the user can rapidly retrieve the stellar mass and age, among many other parameters.

The mass of the targets varies between 0.17 and 0.90 \Msun. Only JW180, JW293, and JW908 are very low-mass stars. Most of the targets have ages around 3 Myr, with the exception of JW847, which is 1 Myr.

From the HR diagram, JW647 and JW847 are the stars that deviate the most from the original classification of \cite{hillenbrand1997}. For the case of JW647, this star seems to be hotter and older. Conversely, JW847 appears to be colder and younger. The remaining stars do not diverge too much on stellar mass, but there are some differences concerning their ages. We will discuss these discrepancies in Sect.\ref{sect:discussion}. 

\section{Accretion analysis}\label{sect:analysys}

The existence of a Balmer jump, veiling and accretion tracers in emission in the spectrum of a YSO give us important information concerning the presence of accretion activity. We explore these points in the following subsections as well as the presence of circumstellar disks through available photometry.

\subsection{Veiling determination}\label{sect:veiling}

When determining the target's spectral type,  we came up against some difficulties with the star JW647. The spectral energy distribution (SED) was slightly different, and the photospheric lines were not as deep as expected. This veiling effect of photospheric lines, very common in classical T Tauri stars, can be explained by an extra continuum source in the UV and visible bands. It is thought that the continuum excess is a consequence of the  shock-heated gas that was transported from the disk to the stellar surface through accretion columns \citep{basri1990,hartmann2016,rei2018}.

In the VIS wavelength band, we extracted the excess emission  by measuring the veiling  of the absorption lines in several spectral ranges from the blue to the red. Before the veiling determination, the targets  spectra were compared with the templates indicated in Table  \ref{table:templates} and broadened to the $v \sin i$ shown in Table \ref{table:1}. The veiling factor was estimated by increasing the value to the broadened template until it matched the target spectrum. Although the resolution is not adequate for an accurate veiling  determination, it is enough to have a rough estimate of the veiling with an error less than 0.2. In Table \ref{table:stellarparams}, we report the veiling measured at 610 nm, which  is negligible for all stars except JW647. For this star, the photospheric stellar flux was rescaled  at wavelength $\lambda$ = 610 nm to the level

\begin{equation}
    F^{\textup{phot}}_{610nm} = \frac{F^*_{610nm}}{r_{610nm} + 1} \, ,
\end{equation}
where r$_{610nm}$ is the veiling factor  and $F^*_{610nm}$ the observed flux measured at this wavelength range.
This rescaling is consistent with the veiling measurements in other wavelength ranges.

\subsection{Balmer continuum and emission lines}

Another example of spectral evidence that the star is accreting is the presence of the Balmer jump near 3646 \AA.
Following \cite{herczeg2008}, the observed Balmer jump ($BJ_\textup{obs}$) is defined as the ratio between the flux at 360 nm and 400 nm. The authors suggested a criterion in which a mid-M dwarf should be considered an accretor if it has a Balmer jump larger than 0.5. This threshold was proposed after the authors measured the $BJ_\textup{obs}$ in a sample of low-mass stars and checked that the non-accreting stars had values systematically below 0.5.

The values measured for the $BJ_\textup{obs}$ are listed in Table \ref{table:stellarparams} for all the targets. According to this criterion, the only star that is clearly accreting is JW647, with a $BJ_\textup{obs}$ of 1.1. The remaining targets have values that match the threshold value and we cannot infer anything about accretion activity for these three cases solely from the observed Balmer jump. Nevertheless, it could be that JW180, JW293, and JW908 are transiting from CTTS (classical T Tauri star) to WTTS (weak-line T Tauri star) evolutionary stage.
JW847 was classified as a K6, and for that reason we should be cautious when using this criterion, since this is not an M-type star. Although this criterion should be valid for mid-M dwarfs, \cite{rugel2018} used this criterion from K6 to M6 type stars belonging to the $\eta$ Chamaeleontis association.

We found several emission lines among the targets' spectra, except for JW847. The remaining YSOs show several Balmer lines in emission, as well as He I lines. The star JW647 shows additional emissions in the NIR arm for the Paschen and Brackett lines. The existence of these emissions is already indicative of ongoing accretion in these YSOs.
Though the low resolution in the current spectra  does not allow an accurate estimate of the  radial velocity of these objects, the spectral emission lines position are consistent with the stellar radial velocity. In Appendix \ref{app:accretion_tracers} we show the profiles of some of the emission lines available for JW180, JW293, JW647, and JW908 (red lines) and compare with the corresponding templates (blue lines).

\begin{figure}
   \centering
   \includegraphics[width=0.8\hsize]{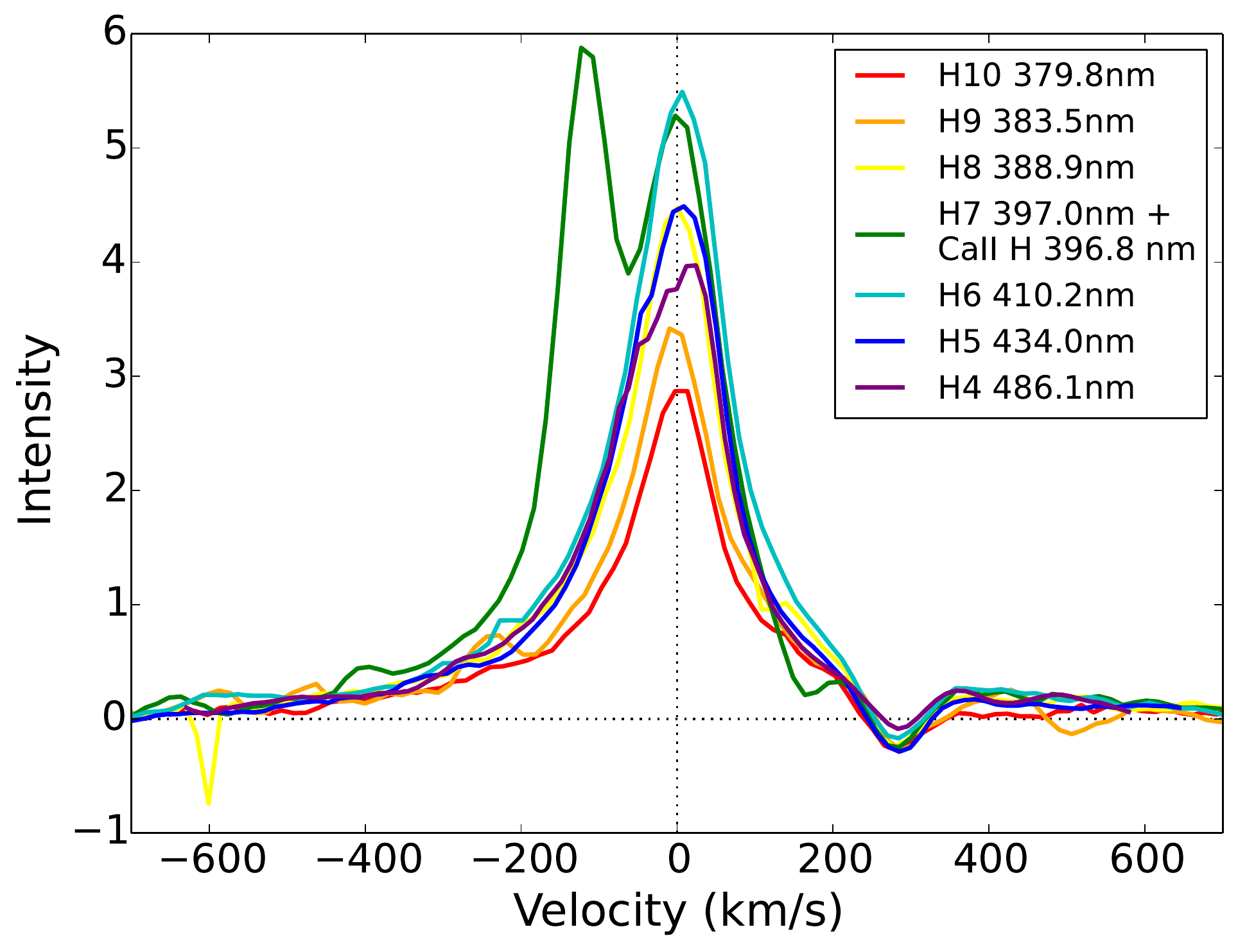}
      \caption{Balmer emission lines in velocity for JW647 with overlapping inverse P Cygni profiles. Each color corresponds to a different Balmer emission line. The H7 line is blended with \ion{Ca}{II} H.}
         \label{fig:Balmer_lines}
   \end{figure}

\begin{figure}
   \centering
   \includegraphics[width=0.8\hsize]{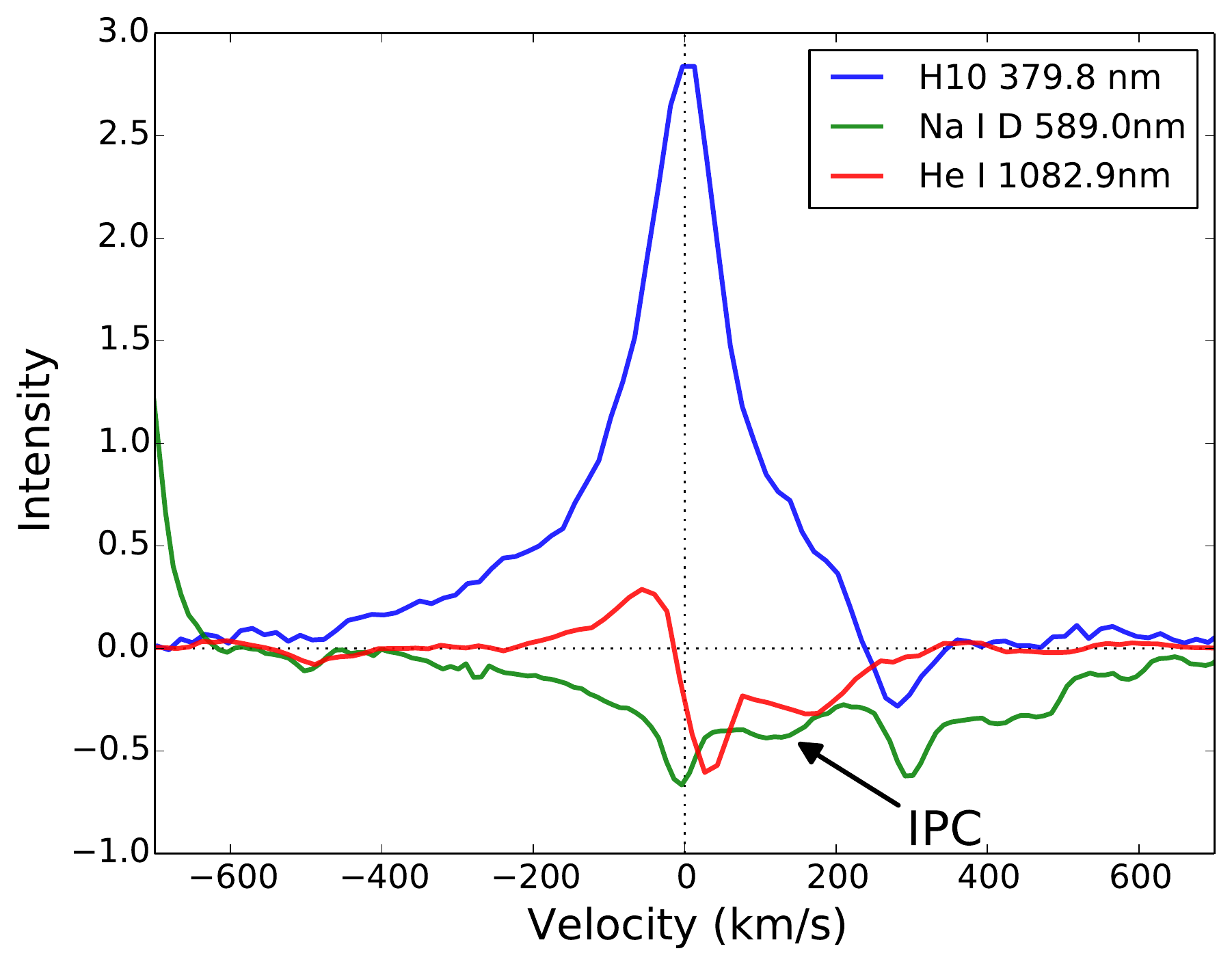}
      \caption{H10, \ion{Na}{I} D and He I emission lines in velocity for JW647 with inverse P Cygni profiles. Each color corresponds to a different line. The IPC for the \ion{Na}{I} D line is marked in the figure.}
         \label{fig:emission_lines_IPC}
   \end{figure}

In the particular case of the JW647 spectrum, the Balmer emission lines from H4 (486.1 nm) to H10 (379.8 nm) show an inverse P Cygni (IPC) absorption present in all the profiles (see Fig.\ref{fig:Balmer_lines}). The red edge of this redshifted absorption falls around 300 \kms, which is the typical value found for infalling material that is accreting onto the star \citep[e.g.,][]{edwards1994}. Such IPC profiles are also observed for the \ion{Na}{I} D and \ion{He}{I} (1082.9 nm) lines, as shown in Fig. \ref{fig:emission_lines_IPC}. The \ion{He}{I} lines at 587.6 and 667.8 do not show IPC profiles. In this last figure, the redshifted absorptions do not overlap, since we would expect emission lines of different elements to be emitted in different regions of the accretion column. Therefore, the corresponding velocities of infalling material should be different.
These redshifted absorptions were first detected in active T Tauri stars by \cite{walker1972}, and they characterize the class of YY Orionis stars. High-velocity redshifted absorptions come from photons emitted near the shock region that are absorbed by the infalling gas in the accretion columns, with high velocity with respect to our line of sight \citep{edwards1994}. Besides the detected veiling, this is another indication that JW647 is an accreting star.

The existence of several accretion tracers in emission, as well as the presence of the Balmer jump near 3646 \AA~in the spectra, suggest that the targets have ongoing accretion activity. Nevertheless, we cannot discard the possibility that these spectral features may also be due to chromospheric activity, especially in the cases of JW180, JW293, and JW908. In order to verify this, we list in Table \ref{table:FWHM} the full width at half maximum (FWHM) measured for some chromospheric activity indicators, namely \ion{Ca}{II} H and K lines and some Balmer lines less contaminated with noise (H$\epsilon$, H$\delta$, H$\gamma$ and H$\beta$). In the case of JW180 and JW293, the FWHM measurements reveal that the emission lines are broader when compared to the ones in the templates. This is not the case for JW908. Four out of six emission lines in the M4.5 template seem to be broader than the ones in JW908. If we compare the same target with an M5 template, all the previous lines of JW908 are broader than the template.
In view of this, among JW180, JW293, and JW908, only the latter is dominated by chromospheric activity rather than accretion activity.

\begin{table*}[]
\centering
\caption{Full widths at half maximum (FWHM) measured for the strongest emission lines, in \kms, for the ONC accreting targets and corresponding templates.}
\label{table:FWHM}
\begin{tabular}{ccccccccc}
\hline\hline
Targets & SpT & \ion{Ca}{II} K & \ion{Ca}{II} H & H7 (H$\epsilon$) & H6 (H$\delta$) & H5 (H$\gamma$) & H4 (H$\beta$) \\
\hline
JW180 & M5 & 41.69 $\pm$ 0.40 & 39.87 $\pm$ 0.24 & 103.75 $\pm$ 2.32 & 90.13 $\pm$ 3.47 & 91.75 $\pm$ 5.15 & 96.93 $\pm$ 2.78 \\
JW293 & M4.5 & 52.41 $\pm$ 0.50 & 51.61 $\pm$ 0.28 & 93.55 $\pm$ 1.32 & 232.10 $\pm$ 280.12 & 92.76 $\pm$ 18.10 & 85.37 $\pm$ 5.45 \\
JW647 & M1 & 29.91 $\pm$ 3.86 & 42.72 $\pm$ 1.44 & 100.69 $\pm$ 6.04 & 148.03 $\pm$ 7.20 & 157.03 $\pm$ 4.84 & 165.83 $\pm$ 8.18 \\
JW908 & M4.5 & 35.14 $\pm$ 0.55 & 33.35 $\pm$ 0.05 & 72.32 $\pm$ 2.44 & 73.99 $\pm$ 3.89 & 78.69 $\pm$ 6.95 & 80.29 $\pm$ 3.86 \\
\hline
Templates &  &  &  &  &  &  & \\
\hline
Par-Lup-3-2 & M5 & 29.33 $\pm$ 1.26 & 27.94 $\pm$ 0.43 & 61.80 $\pm$ 1.69 & 65.87 $\pm$ 3.02 & 71.41 $\pm$ 6.86 & 72.50 $\pm$ 3.21 \\
SO797 & M4.5 & 36.55 $\pm$ 1.89 & 35.21 $\pm$ 0.75 & 76.34 $\pm$ 2.11 & 77.57 $\pm$ 4.33 & 78.39 $\pm$ 5.12 & 77.72 $\pm$ 1.49 \\
TWA13B & M1 & 20.64 $\pm$ 0.34 & 19.68 $\pm$ 0.03 & 49.53 $\pm$ 1.58 & 47.46 $\pm$ 4.43 & 49.22 $\pm$ 4.60 & 59.56 $\pm$ 6.57 \\
\hline
\end{tabular}
\end{table*}

\subsection{Mass accretion rates}

One way to quantify accretion in YSOs is through the measurement of line fluxes ($f_\textup{line}$), which can be converted to line luminosities ($L_\textup{line}$), and corresponding accretion luminosities ($L_\textup{acc}$) (e.g., \cite{alcala2014}). Once we know this later parameter, plus the stellar mass ($M_*$) and radius ($R_*$), we can estimate the mass accretion rates (\Macc).

In order to derive the mass accretion rates, we analyzed several accretion indicators from the UVB to the NIR arms. The emission lines used in this estimate are listed in Table \ref{table:6}. JW647 is the only target that shows Pa$\beta$ and Br$\gamma$ NIR lines in emission, which can be found in actively accreting stars. Although the emission line of H$\alpha$ is a well-known accretion tracer, we discarded this line to measure the mass accretion rates because it is very likely contaminated with nebular emission \citep{muench2008}.

In a first step, we measured the line fluxes of all the previous lines, and we converted them to line luminosities by assuming the stellar distances listed in Table \ref{table:1}. The accretion luminosity was estimated through the empirical relation 
\begin{equation}
    \log (L_\textup{acc} / L_\textup{\sun}) = a \log(L_\textup{line} / L_\textup{\sun}) + b \, ,
\end{equation}
with the coefficients $a$ and $b$ from \cite{alcala2014}. 
From the accretion luminosities, we can now derive the mass accretion rates with the following equation:
\begin{equation}
    M_\textup{acc} = \left ( 1 - \frac{R_*}{R_\textup{in}} \right )^{-1} \frac{L_\textup{acc} R_*}{G M_*} \approx 1.25 \frac{L_\textup{acc} R_*}{G M_*} \, ,
\end{equation}
where $R_\textup{in}$ is the star's inner-disk radius, which corresponds to the truncation radius of the disk. Like in \cite{gullbring1998} and \cite{alcala2014}, we assume $R_\textup{in} = 5R_*$. The measured \Macc ~for the different lines are listed in Table \ref{table:6}. Since the emission lines in JW908 are likely chromospheric lines, we cannot use them to derive the corresponding mass accretion rates. For this reason, in Table \ref{table:6} we list only the line fluxes for this target. According to the previous table, JW647 is the star with the highest average accretion rate, with a value of 10$^{-8.7} \, M_\odot yr^{-1}$. Since we had only very few and very weak accretion tracers in emission for JW847, we could not estimate the accretion rate for this star. Additionally, we suspect that the \ion{Ca}{II} emission lines in the UVB arm have included some chromospheric emission that could affect the \Macc~ determination. The remaining targets had ten accretion tracers in emission that returned a \Macc~around 10$^{-10.1} \, M_\odot yr^{-1}$.
According to \cite{hartmann2016}, the estimated mass accretion rates are within the expected values for this range of stellar masses.

\begin{sidewaystable*}
\caption{Line fluxes measured for the accretion/chromospheric  tracers in erg s$^{-1}$ cm$^{-2}$ and the logarithmic mass accretion rates derived for the ONC targets in $M_\sun yr^{-1}$.}
\label{table:6}     
\centering                         
\begin{tabular}{lllllllll}       
\hline\hline
Line & $\lambda$ (nm) & \multicolumn{2}{c}{JW180} & \multicolumn{2}{c}{JW293} & \multicolumn{2}{c}{JW647} & JW908 \\
 &  & $f_\textup{line}$ & \Macc & $f_\textup{line}$ & \Macc & $f_\textup{line}$ & \Macc & $f_\textup{line}$\\
\hline
H10 & 379.8 & 2.39($\pm$0.40)e-16 & -10.1$\pm$0.2 & 1.70($\pm$0.26)e-16 & -10.2$\pm$0.1 & 4.94($\pm$0.20)e-15 & -8.8$\pm$0.1$^*$ & 1.68($\pm$0.47)e-16 \\
H9 & 383.5 & 2.71($\pm$0.33)e-16 & -10.2$\pm$0.2 & 1.70($\pm$0.12)e-16 & -10.3$\pm$0.1 & 5.94($\pm$0.38)e-15 & -8.8$\pm$0.1$^*$ & 2.11($\pm$0.36)e-16 \\
H8 & 388.9 & 3.75($\pm$1.04)e-16 & -10.2$\pm$0.2 & 2.80($\pm$0.23)e-16 & -10.2$\pm$0.1 & 7.63($\pm$0.37)e-15 & -8.8$\pm$0.1$^*$ & 2.30($\pm$0.22)e-16 \\
\ion{Ca}{II} (K) & 393.4 & 8.74($\pm$0.13)e-16 & -9.8$\pm$0.2 & 7.59($\pm$0.12)e-16 & -9.8$\pm$0.1 & 3.83($\pm$0.33)e-15 & -9.2$\pm$0.1$^*$ & 9.37($\pm$0.22)e-16 \\
\ion{Ca}{II} (H) & 396.8 & 7.27($\pm$0.09)e-16 & -9.9$\pm$0.2 & 7.02($\pm$0.07)e-16 & -9.9$\pm$0.1 & 5.99($\pm$0.19)e-15 & -9.0$\pm$0.1 & 7.84($\pm$0.48)e-16 \\
H7 (H$\epsilon$) & 397.0 & 3.64($\pm$0.16)e-16 & -10.3$\pm$0.2 & 2.44($\pm$0.06)e-16 & -10.4$\pm$0.1 & 5.89($\pm$0.93)e-15 & -8.9$\pm$0.1$^*$ & 2.46($\pm$0.13)e-16 \\
\ion{He}{I} & 402.6 & ... & ... & ... & ... & 7.65($\pm$0.17)e-16 & -8.8$\pm$0.1 & ... \\
H6 (H$\delta$) & 410.2 & 3.92($\pm$0.23)e-16 & -10.3$\pm$0.2 & 4.69($\pm$0.40)e-16 & -10.2$\pm$0.1 & 1.21($\pm$0.07)e-14 & -8.7$\pm$0.1$^*$ & 3.17($\pm$0.25)e-16 \\
H5 (H$\gamma$) & 434.0 & 6.02($\pm$0.49)e-16 & -10.3$\pm$0.2 & 6.15($\pm$0.85)e-16 & -10.2$\pm$0.1 & 1.46($\pm$0.05)e-14 & -8.7$\pm$0.1$^*$ & 5.13($\pm$0.62)e-16 \\
\ion{He}{I} & 447.1 & ... & ... & ... & ... & 1.29($\pm$0.10)e-15 & -8.7$\pm$0.1 & ... \\
H4 (H$\beta$) & 486.1 & 1.18($\pm$0.05)e-15 & -10.3$\pm$0.2 & 1.45($\pm$0.08)e-15 & -10.1$\pm$0.1 & 1.85($\pm$0.12)e-14 & -8.9$\pm$0.1$^*$ & 1.09($\pm$0.07)e-15 \\
\ion{He}{I} & 587.6 & 1.65($\pm$0.20)e-16 & -10.2$\pm$0.2 & 1.97($\pm$0.31)e-16 & -10.0$\pm$0.1 & 3.75($\pm$0.63)e-15 & -8.6$\pm$0.1 & 1.43($\pm$0.40)e-16 \\
\ion{He}{I} & 667.8 & ... & ... & ... & ... & 1.15($\pm$0.12)e-15 & -8.7$\pm$0.1 & ... \\
\ion{He}{I} & 706.5 & ... & ... & ... & ... & 8.43($\pm$0.96)e-15 & -8.7$\pm$0.1 & ... \\
Pa7 (Pa$\delta$) & 1004.9 & ... & ... & ... & ... & 7.46($\pm$1.87)e-15 & -8.6$\pm$0.2 & ... \\
Pa6 (Pa$\gamma$) & 1093.8 & ... & ... & ... & ... & 1.21($\pm$0.23)e-14 & -8.5$\pm$0.1 & ... \\
Pa5 (Pa$\beta$) & 1281.8 & ... & ... & ... & ... & 8.27($\pm$1.27)e-15 & -8.9$\pm$0.1 & ... \\
Br7 (Br$\gamma$) & 2166.1 & ... & ... & ... & ... & 3.05($\pm$0.12)e-15 & -8.7$\pm$0.1 & ... \\
\hline
Mean &  &  & -10.1 &  & -10.1 &  & -8.7 &  \\
\hline                         
\end{tabular}
\tablefoot{(*) This emission line shows an Inverse P Cygni (IPC) profile.}
\end{sidewaystable*}

\subsection{Evidence of circumstellar disks}

One way to infer the presence of circumstellar disks among the selected YSOs, is to look for infrared excess emission through the SED of these objects. Usually, class II objects have infrared excess emission revealed through energy distributions with broader profiles, when compared with the corresponding black-body SED. After 2$\mu$m, the energy distribution can show flat or negative slopes \citep{lada1987}. For class III objects, these have SEDs characteristic of a stellar photosphere \citep{whitney2003}.

\begin{figure*}
\centering
\begin{tabular}{cc}
  \includegraphics[width=0.45\linewidth]{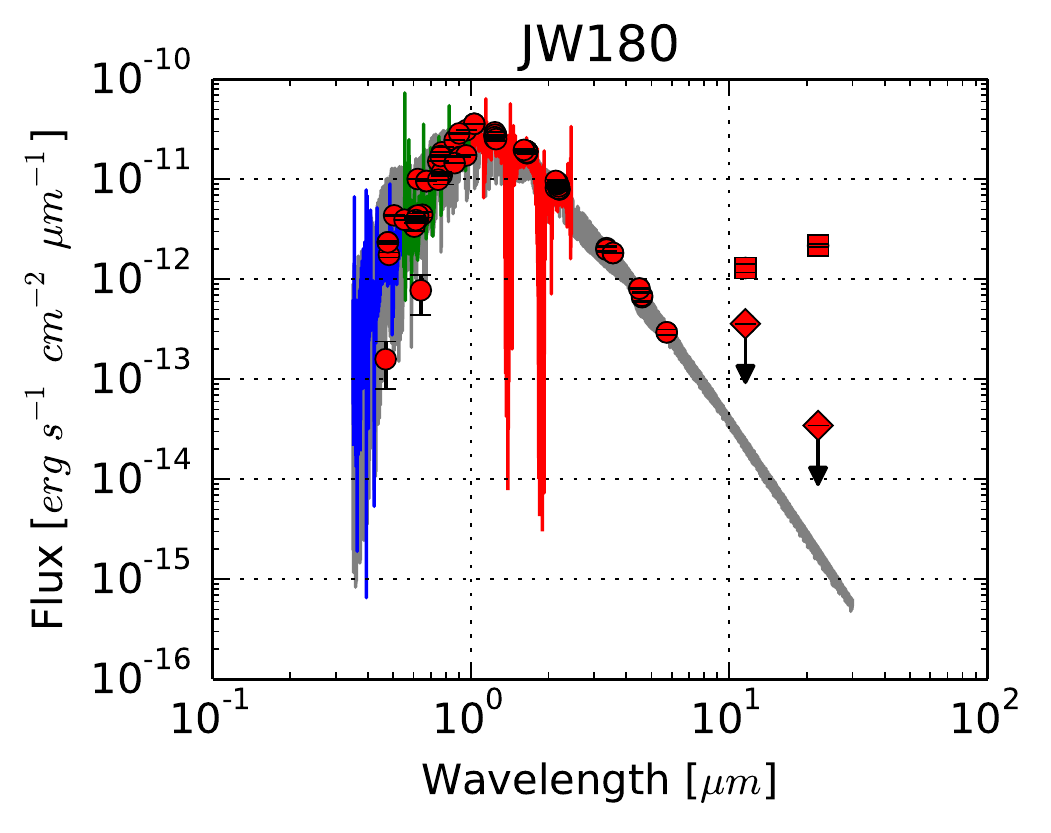} &   \includegraphics[width=0.45\linewidth]{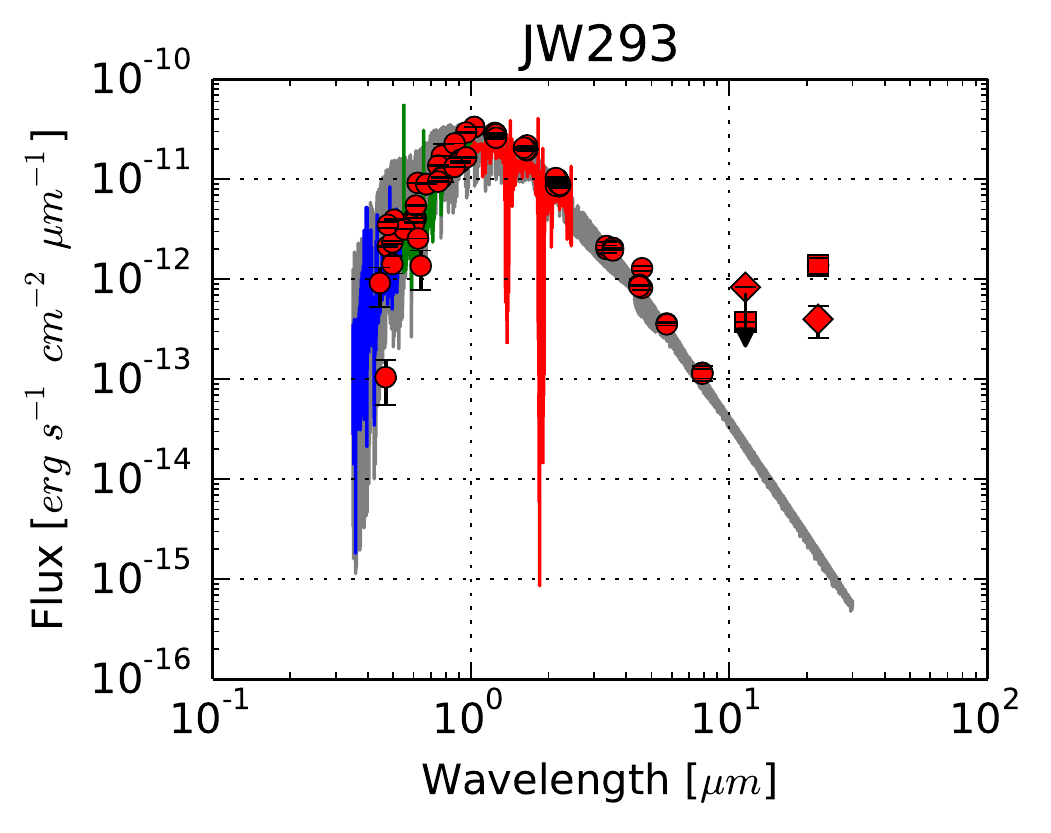} \\
(a) & (b) \\[6pt]
 \includegraphics[width=0.45\linewidth]{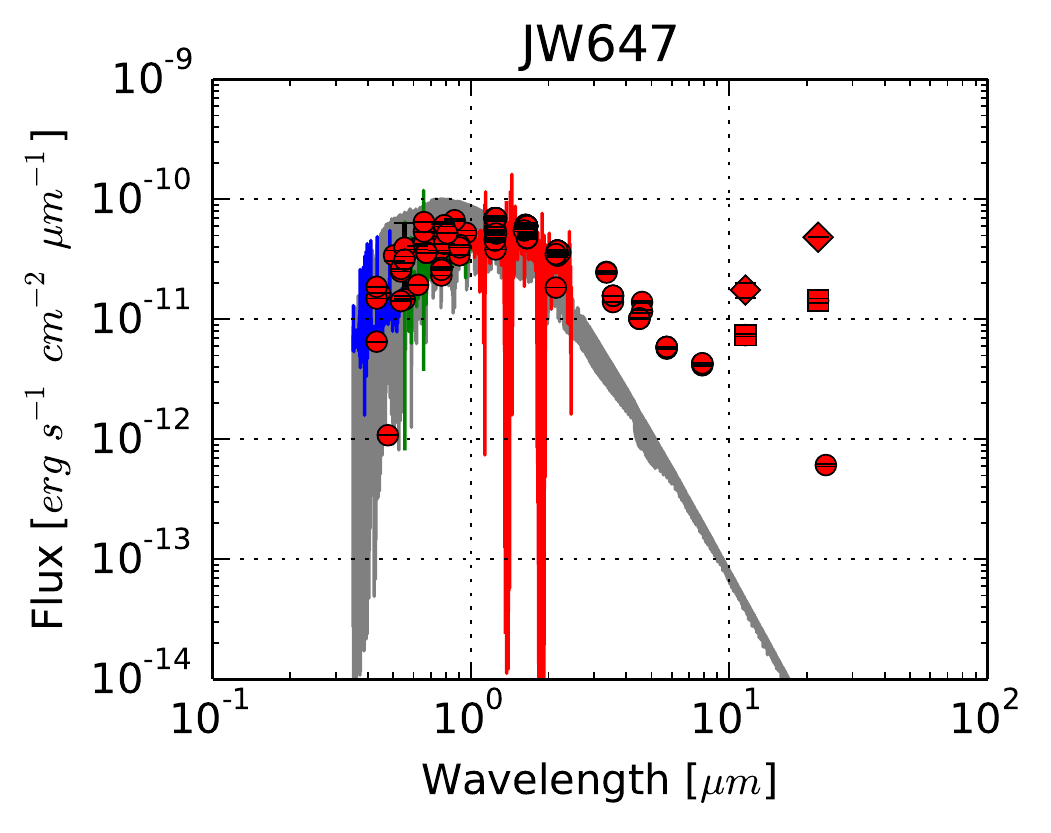} &   \includegraphics[width=0.45\linewidth]{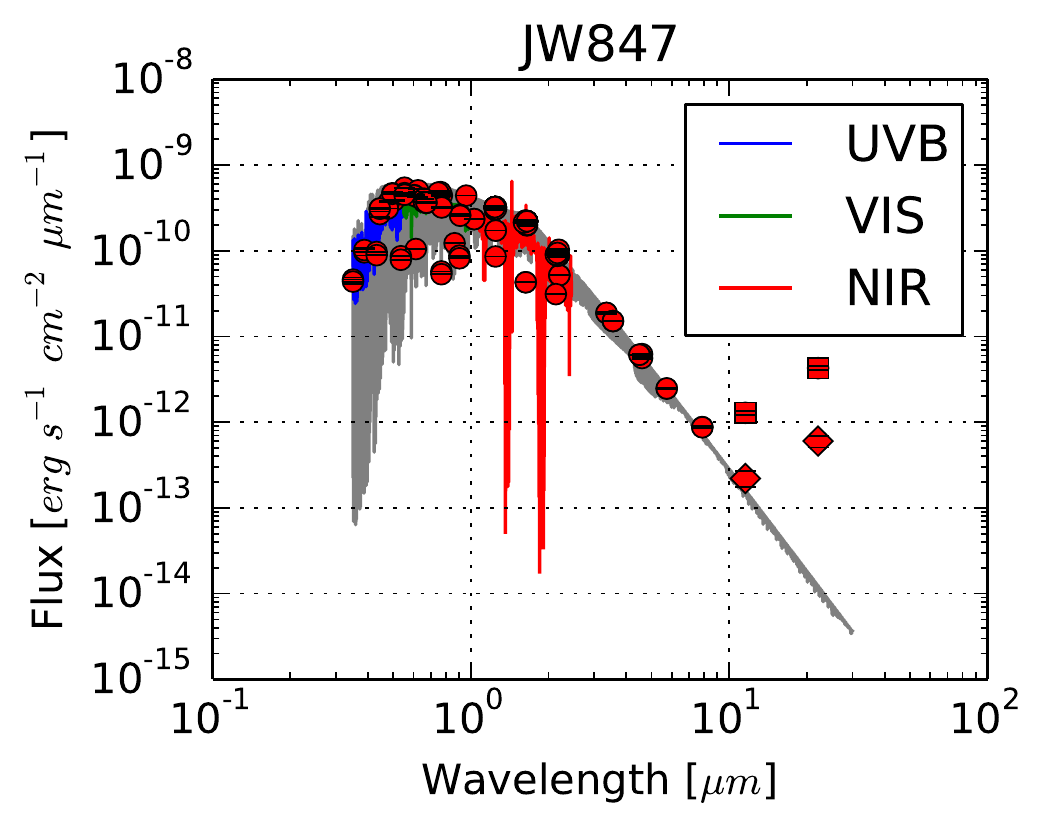} \\
(c) & (d) \\[6pt]
\multicolumn{2}{c}{\includegraphics[width=0.45\linewidth]{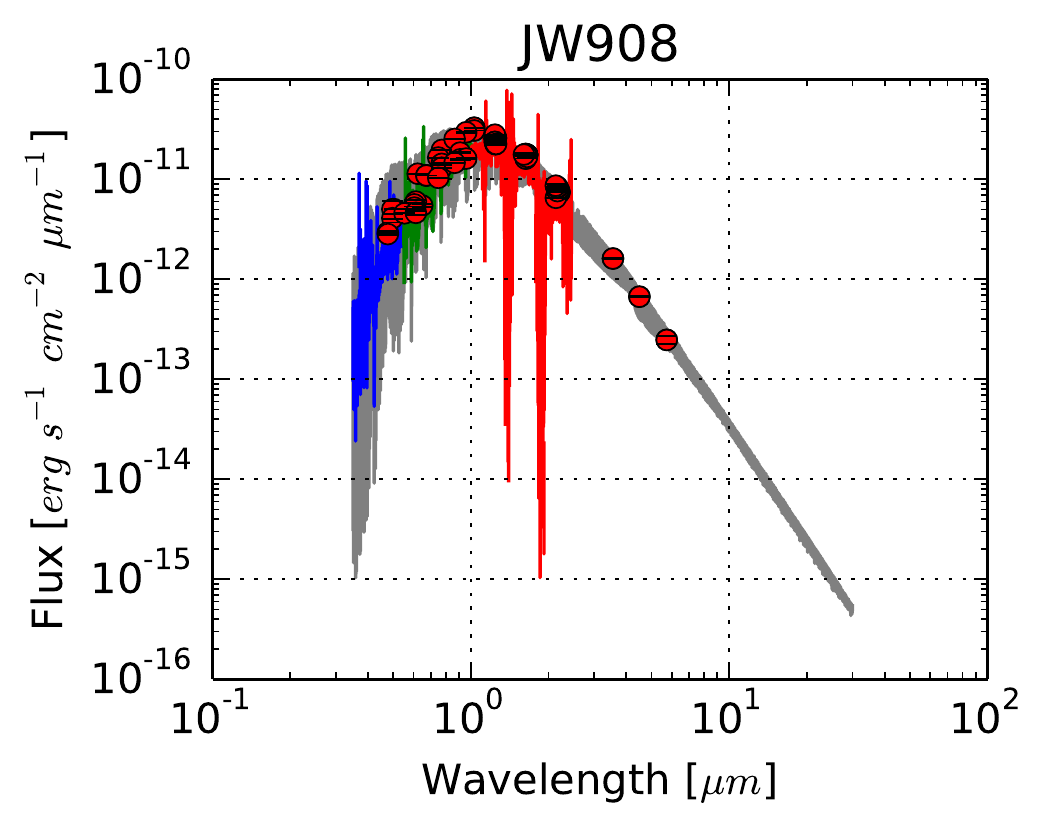}}\\
\multicolumn{2}{c}{(e)}
\end{tabular}
\caption{Spectral energy distributions of targets. The spectra of the stars are shown for different wavebands: ultra-violet (blue), visible (green), and infrared (red). The gray line corresponds to the BT-Settl synthetic spectra from \citet{allard2012} with effective temperatures of 3100, 3200, 3700, 4200, and 3200 K (for plots a, b, c, d, and e, respectively), $\log g = 4.0$, solar metallicity and zero chemical index. The red dots correspond to the available photometry at Vizier \citep{ochsenbein2000} for a search radius of 0.5\arcsec. The squares and diamonds correspond to WISE \citep{cutri2012} and AllWISE \citep{cutri2013} photometry, respectively, for the W3 and W4 bands. The black arrows represent upper limits for the photometric values shown.}
\label{fig:SEDs_BT-Settl_05arcsec}
\end{figure*}

In Fig. \ref{fig:SEDs_BT-Settl_05arcsec}, we compare the SED of each YSO with synthetic spectrum and photometric data. The gray line corresponds to the synthetic BT-Settl spectra \citep{allard2012}, with the closest temperature to the one we derived in this work for each target. The red dots correspond to photometric data, for a search radius of 0.5 \arcsec, available at Vizier \citep{ochsenbein2000}, which gathers data from several surveys, namely, Gaia DR1 and DR2 \citep{gaia2016,gaia2018}, 2MASS \citep{cutri2003}, WISE \citep{cutri2012}, AllWISE \citep{cutri2013}, Pan-STARRS \citep{chambers2016}, SkyMapper Southern Sky Survey \citep{wolf2018}, and VISTA Hemisphere Survey \citep{mcmahon2013}. The blue, green, and red solid lines correspond to the X-shooter spectra in the ultraviolet, visible, and near-infrared, respectively. The synthetic spectra were scaled according to the J-band value ($\sim1.25 \, \mu m$) of the star.
With the exception of JW647 and JW847, the photometric data and flux calibrated spectra are in good agreement. For those two exceptions, we observe a larger spread on photometric values.

According to the \cite{lada1987} classification, JW908 resembles the SED of a class III object, but this is not clear, since we do not have photometry beyond 10 $\mu m$. The remaining YSOs show an infrared excess characteristic of class II objects. The plots also show a dip slightly before $10 \, \mu m$ for JW180, JW293, JW647, and JW847. According to \cite{furlan2009}, there is silicate emission feature from the disk, which appears at $10 \, \mu m$. The equivalent width of this emission can be used to identify objects whose disks start to develop gaps.

Although we do not have infrared spectroscopy beyond 2.5 $\mu m$, from the SED profiles of JW180, JW293, JW647, and JW847, we suspect that we might have some transition disks among them (e.g., \cite{williams2011} and references therein), as we explain ahead in this subsection. These circumstellar disks whose inner regions were partially cleared from dust can be related to planet formation or disk dispersal around pre-main sequence stars \citep{espaillat2014}.

 \cite{strom1989} and \cite{skrutskie1990} were the pioneers of transition disk detection through NIR and MIR (mid-infrared) photometry. These objects show a small excess emission at NIR wavelengths (1-5$\mu m$) in their SEDs, followed by an excess at the MIR (5-20$\mu m$). This can be understood as the presence of a hole in the disk without dust, scarcity of dust particles in the inner disk, or even a change in the dust particle size (see review by \cite{carmona2010} and references therein). Furthermore, it has been suggested that these disks result from the transition from Class II to Class III objects, where optically thick disks evolve towards optically thin ones with progressive dissipation of its gas and dust.

\cite{espaillat2014} distinguished three types of disks: a full disk, a pre-transitional disk, and a transitional disk. A full disk corresponds to a continuous disk without significant gaps. A transitional disk corresponds to a disk with an inner disk hole, while a pre-transitional disk has a gap that separates an inner dust ring from an outer one. 

When comparing the SED from a full disk with a pre-transitional disk or a transitional disk, the dip on the left of $10 \, \mu m$ silicate feature gets deeper as the warm dust in the inner region of the disk starts do disappear.
One of the main differences of the SEDs between the pre-transitional and transitional phase is the significant NIR excess that pre-transitional disks show comparatively to their corresponding stellar photospheres, which is not as pronounced for transitional disks. This can be explained by the presence of an optically thick inner disk close to the star, which corresponds to the residuals of the inner portion of the original disk. Once this inner ring starts to dissipate, there is no NIR excess in the SED.

Looking at the four top plots in Fig. \ref{fig:SEDs_BT-Settl_05arcsec}, we can see that JW647 shows a pronounced NIR excess emission compared to JW180, JW293, and JW847. Taking into account the \cite{espaillat2014} classification and the SEDs of these four objects, we suggest that JW647 has a pre-transitional disk, while JW180, JW293, and JW847 have a transitional disk. Concerning JW908, since we do not have MIR photometry available, we cannot infer if this star hosts a transitional disk or not.

There is, at least, one work in the literature mentioning the detection of transition disks in ONC members \citep{furesz2008}. Those authors analyzed a  large sample of stars through \textit{Spitzer} photometry and confirmed that 35 objects have transition disks. 
In this work, JW293 is confirmed to have a transition disk, while this is not certain for JW908. Concerning JW647, although this star shows clear signs of accretion, the disk should not have started cleaning its inner orbits yet.
As mentioned by the authors, the transition disks identified in the ONC should be among the youngest systems known with only few Myr. In our case, the objects have ages ranging between 1 and 3.5 Myr.

\section{Discussion}\label{sect:discussion}

The study of accretion activity among the very low-mass members of the ONC has not been fully addressed. Although this work has quite a small sample, here we make the effort to get a step closer toward the bigger picture. \cite{biazzo2009} studied the disk-locking scenario among low-mass stars of the ONC and concluded the percentage of accretors seems to scale inversely to the mass of the stars. By using the H$\alpha$ width at 10\%, they did not find any accreting stars with masses smaller than 0.2 M$_\odot$, but they found that some of those stars were locked to their disks as they showed slow rotation periods and infrared excess. 
The evidence that the YSOs studied in \cite{biazzo2009} have, apparently, no ongoing accretion, but they seem to be locked to their disks, suggests that disk-locking took place extremely quickly and early in their evolution. If this is confirmed, it may bring implications for the timescale of the angular momentum evolution in low-mass stars.

A small sample of possible accreting stars was selected, and six out of eight were observed. Additional problems concerning the observation mode and saturation issues decreased the sample down to the five objects presented in this study. It turned out that only three of those five targets are very low-mass stars. But, do they have ongoing accretion?
In order to disentangle all the findings, in the following subsections, we discuss some of the main results found for each target.

\subsection{JW180}
JW180 was classified as an M5 and shows the lowest luminosity, stellar mass, and radius computed, and it has an estimated age of 3.5 Myr. This star has quite a few accretion tracers in emission, most of them being Balmer lines, which were translated in a \Macc~of $10^{-10.1} M_\sun yr^{-1}$. According to the corresponding SED, there is a significant MIR excess indicative of the presence of a disk whose inner region has been cleared. 
This MIR photometry was taken from the WISE and AllWISE catalogs and corresponds to bands W3 ($12 \, \mu m$) and W4 ($22 \, \mu m$) with an angular resolution of 6.5\arcsec and 12\arcsec, respectively. These two bands are represented in Fig. \ref{fig:SEDs_BT-Settl_05arcsec} with red squares and diamonds for WISE and AllWISE photometry, correspondingly. The AllWISE Source Catalog is an improved version of the WISE All-Sky Release Catalog, and for such matters, the AllWISE photometry should be more reliable. In order to check if this photometry is trustworthy, we can take a look at the quality flag parameters available in the AllWISE catalog, namely the photometric quality flag (qph) and the contamination and confusion flag (ccf), for both W3 and W4 bands. According to the ccf parameter, we confirm that we have targets with bands flagged as spurious detections of images artifacts, which means that we should be cautious about the reliability of the source.

From the qph parameter, we confirm that, for JW180, the photometric values for these two bands are flagged as upper limits, which are represented by the black arrows in Fig. \ref{fig:SEDs_BT-Settl_05arcsec}. When we look to the ccf parameter, we see that, for the band W3, the photometric value may be a spurious detection or it be contaminated by a scattered light halo surrounding a nearby bright source, which exists at approximately 30\arcsec~from JW180. Concerning W4, this band is flagged with a diffraction spike, which could be from a nearby bright star on the same image. We checked the corresponding image at NASA and the IPAC Infrared Science Archive and there is a particular bright star, HD3698, that could cause this effect.

Taking everything into account, we cannot infer solely from WISE and AllWISE photometry if this star is hosting a transitional disk, since the MIR photometry is flagged with upper limits. Nevertheless, given the several accretion tracers in emission for this star and the computed mass accretion rate for a star of 0.17\Msun, it is very likely that this object would still be surrounded by a disk.

\subsection{JW293 and JW908}
These two objects were classified with the same spectral type as M4.5. They have very similar luminosity, stellar mass, radius, and their age is around 3 Myr. JW293 has the same lines in emission as JW180, returning an average \Macc~of $10^{-10.1} M_\sun yr^{-1}$. The SED of JW293 has a similar MIR excess to JW180. 

We checked that we have another upper limit marked in the MIR photometry of JW293, but only concerns the W3 band. This same band is flagged with an "optical ghost" in the ccf parameter, which could be caused by a nearby bright source. This could be due either to proximity to the nebula, which is causing background contamination, or to the presence of the bright star HD 36981 in the corresponding image from IPAC.

Unfortunately, we do not have photometry in the MIR for JW908. As it is, the SED resembles a class III object. Given all the accretion tracers in emission for this star, we could think that JW908 is accreting. However, we saw previously that this ONC member is likely dominated by chromospheric activity. This could suggest that JW908 is transiting from CTTS to WTTS.

These results agree with the classification given by \cite{furesz2008}, who classified JW293 as: ``probably a CTTS, but could be a WTTS'', and JW908 as a WTTS. In principle, we conclude that JW180 has ongoing accretion activity. Concerning JW908, it is not so clear with the current data.

\subsection{JW647}

JW647, originally classified as an M5 in \cite{hillenbrand1997}, was recently changed to an M0.5 in \cite{hillenbrand2013}. According to our work, we suggest a spectral type of M1 for this YSO. This is what explains a big shift in effective temperature in the HR diagram. Concerning the shift in luminosity, the flux for JW647 was corrected from veiling in this work, while in \cite{hillenbrand1997} it was not possible to correct the spectral types from this additional contribution. That should be the reason why the luminosity drops slightly in the same diagram. Comparing with the previous targets, JW647 has a higher stellar mass of approximately 0.5 \Msun \, and an age of about 3 Myr. However, this age estimate is highly dependent on the flux correction due to the continuum excess emission determined through the veiling of photospheric lines.

This was the only star in the sample that showed significant veiling, with an average value of 0.5 measured at 610 nm. A strong Balmer jump was also observed in the UVB arm, comparatively to the remaining targets. This characteristic reflects itself on the accretion criterion from \cite{herczeg2008} with a measured observed Balmer jump of 1.1, the highest in the sample. Additionally, we also detected several accretion tracers for this YSO, and many emission lines present inverse P Cygni profiles. Taking all the emissions into account, for this star we got the highest \Macc~with a value of $10^{-8.7} M_\sun yr^{-1}$, which is a typical mass accretion rate  for CTTS \citep{hartmann2016}. This result is in agreement with the work of \cite{furesz2008}, who classified this star as a CTTS.

\cite{manara2012} previously determined the mass accretion rates for JW647 (OM 1197) with Hubble Space Telescope observations using U-band excess. The associated error with U-band measurements is $\leq 0.3$ mag. Their value corresponds to $10^{-6.06} M_\sun yr^{-1}$, which is quite high for a star with a stellar mass of 0.47 \Msun. It could be that the \Macc~estimated through photometry is not that precise, but could also be due to the stellar parameters assumed or contamination effects.

The SED of this object stands out from all the others, because we have an additional and significant NIR excess. This excess suggests the presence of a thick inner ring close to the star. This means that JW647 is going through an earlier evolutionary stage compared to the remaining YSOs. This idea can be supported by the mass accretion rate, which is the highest in the sample. Once the YSO is closer to turning into a class III object, accretion rates decrease as the star cleans its disk and there is less and less material to be accreted. 

Finally, no upper limits are signed for the MIR photometry in the AllWISE catalog. Although JW647 has quite close neighbors, at an angular distance of less than 10\arcsec, we can verify that they are not brighter than the target.

\subsection{JW847}
Previously, \cite{hillenbrand1997} classified JW847 as a K3/G8. According to our work, JW847 is the earliest-type star in the sample with a spectral type of a K6, which explains such a big temperature shift in the HR diagram.
This is supported by template comparison, in which the target spectrum best fits a late K-type template rather than an early K-type one.

Besides being the most luminous object with an age of only 1 Myr, it also presents the highest value for stellar mass and radius. The spectrum has no significant accretion tracers in emission, and there were signs of chromospheric emission when we analyzed the \ion{Ca}{II} K and H lines at 393.4 and 396.8 nm, respectively. These line profiles show a narrow emission feature embedded in the line absorption.
Similarly to JW647, \cite{manara2012} estimated the accretion rate for JW847 (OM 2966) with a value of $10^{-7.98} M_\sun yr^{-1}$. This result is very doubtful since there is no evidence in the spectrum of JW847 that this star is an accretor.

We think that this star has no ongoing accretion, but the corresponding SED indicates that there is still an outer disk. If true, this star hosts the youngest transition disk of the sample with only 1 Myr.
Although there are no upper limits flagged for this star according to AllWISE MIR photometry, there is a “halo” flagged for the W3 band that could be due to the presence of a bright star, NV Ori, approximately 30\arcsec~from the target.

It could be that JW847 is already transiting from class II to class III. 
\cite{hernandez2008} plotted the disk frequency of late-type stars (mid-K and later) as a function of age, through NIR excess emission from the disk for several stellar groups. According to their results, at 1 Myr, 80\% of the stars have disks, but 20\% do not. Some stars loose their disks quite early, and this may happen for several reasons, namely the proximity of high-mass stars, binary companions, or even strong interactions with other stars in early star formation.

\section{Conclusions}\label{sect:conclusions}

We analyzed X-shooter spectra of low and very low-mass YSOs belonging to the ONC. These objects were suspected of carrying accretion activity in a previous work from \cite{biazzo2009}. The main goal of this study was to infer if those objects are accretors or not. Firstly, we classified the spectral type of the stars through spectral indices and determined the corresponding stellar parameters. Secondly, we looked for accretion features in the spectra and measured accretion rates in all three arms (UVB, VIS, and NIR) of the suspicious accretors. Finally, we looked for the presence of circumstellar disks through available photometry.

We conclude that JW647 is a CTTS and has ongoing accretion activity supported by its mass accretion rate, IPC profiles present among Balmer emission lines and measured Balmer jump. The SED for this star has a considerable NIR excess that suggests the presence of a thick inner disk as is found in stars with pre-transitional disks.

JW180, JW293, and JW847 show photometric evidence of hosting a transitional disk. 
However, AllWISE MIR photometry is flagged with upper limits for JW180 and at the moment we do not have any other photometry source to clarify if this star really hosts a transition disk or not.
JW180 and JW293 stars are quite similar to each other, not only among the computed stellar parameters but also concerning the measured mass accretion rates. The target JW847 is a different case. 
This K-type star appears to be brighter, bigger, and more massive when compared with JW180 and JW293. 
Nevertheless, it has very few accretion tracers in emission.
It is most likely that JW847 is not accreting and could be transiting from class II to class III. 
While the remaining targets host 3-Myr-old disks, the one from JW847 is only 1 Myr old.

Although we do not have enough photometry to state that JW908 has a circumstellar disk, this star has quite a few accretion tracers in its spectrum that support ongoing accretion. However, there is spectral evidence that chromospheric activity is dominant over accretion activity. Therefore, we cannot discard the possibility that JW908 is transiting from the CTTS to WTTS evolutionary stage.

Taking everything into account, we detect accretion in three out of the five targets with \Macc~values that are in agreement for the corresponding stellar masses \citep{hartmann2016}. Among the three accretors, two of them belong the the very low-mass range with stellar masses below 0.25 \Msun.

Although we cannot retrieve a robust conclusion from only a few YSOs, it seems that the accretion criterion of \cite{white2003} does not work for stars in the very low-mass range. According to this work, three of the targets show spectral evidence of ongoing accretion that was not detected by the previous criterion in \cite{biazzo2009}. A new accretion threshold for such low-mass YSOs should be studied in future work. 
In addition, if disk-locking is occurring, it means that it should happen in an even earlier stage for very low-mass stars. 
A more quantitative analysis would be useful in order to confirm if either disk-locking or any other angular momentum evolution model would fit within the timescale observed for the ONC members. Additionally, we would like to characterize the transition disks of these YSOs more deeply with data from instruments that can give us information not only about the disk mass (IRAM/NOEMA), but also to understand the disk structure (SPHERE), as well as with magnetic topology to understand the star-disk interaction (SPIRou).

\begin{acknowledgements}

This work was supported by Funda\c{c}\~ao para a Ci\^encia e a Tecnologia (FCT) through national funds and by Fundo Europeu de Desenvolvimento Regional (FEDER) through COMPETE2020 - Programa Operacional Competitividade e Internacionaliza\c{c}\~ao by these grants: UIDB/04434/2020; UIDP/04434/2020; UID/FIS/04434/2019; PTDC/FIS-AST/32113/2017 \& POCI-01-0145-FEDER-032113.
RMGA acknowledges support from FCT through the Fellowship PD/BD/113745/2015 (PhD::SPACE Doctoral Network PD/00040/2012) and POCH/FSE (EC) and from Centro de Astrofísica da Universidade do Porto through the research fellowship CIAAUP-23/2019-BIM.
SHPA acknowledges financial support from CAPES, CNPq and Fapemig.
CS thanks LUPM for hosting him during his "délégation CNRS". 
We acknowledge financial support from Programme National de Physique Stellaire (PNPS) of CNRS/INSU (France) and from CRUP through the PAUILF cooperation program (TC-16/17).
We would like to thank J.~M. Alcal\'a and A. Carmona for the enlightening discussions and suggestions, as well as C.~F. Manara and G. F\H{u}r\'esz for confirming some details in the few targets analyzed in common.
This work made use of PyAstronomy and the "Aladin sky atlas" developed at CDS, Strasbourg Observatory, France.
This work has also made use of data from the European Space Agency (ESA) mission {\it Gaia} (\url{https://www.cosmos.esa.int/gaia}), processed by the {\it Gaia} Data Processing and Analysis Consortium (DPAC,
\url{https://www.cosmos.esa.int/web/gaia/dpac/consortium}).
\end{acknowledgements}

\bibliographystyle{aa}
\bibliography{ms}

\begin{appendix}

\section{Impact of the distance in the stellar parameters of JW180}\label{app:distances}

As mentioned in Sect. \ref{sect:sample}, the distances derived from Gaia DR2 parallaxes appear to be smaller relative to average distance values of the ONC, specially for JW180. Some of the Gaia DR2 source parameters were checked in order to see if the parallaxes of the survey are trustworthy, namely the goodness-of-fit statistics of the astrometric solution for the source in the along-scan direction (\textit{astrometric\_gof\_al}), the astrometric goodness of fit $\chi^2$ in the along-scan direction (\textit{astrometric\_chi2\_al}), and the excess noise of the source (\textit{astrometric\_excess\_noise}). According to these values listed in Table \ref{table:GaiaDR2_parameters}, the distance for JW180 is the less reliable one, since the goodness of fit (GOF), the $\chi^2$ value, and the excess noise associated with the parallax measurements reach quite high and undesirable values. These parameters justify why we get an error of approximately 35 pc for this star, the highest of the sample. For the remaining targets, the parameters are lower, leading to errors below 15 pc.

\begin{table}[]
\centering
\caption{Gaia DR2 source parameter values for the goodness-of-fit statistics of the astrometric solution for the source (GOF) and the astrometric goodness of fit $\chi^2$ in the along-scan direction, and the excess noise of the source. The distances derived from the corresponding parallaxes are listed with the corresponding errors.}
\label{table:GaiaDR2_parameters}
\begin{tabular}{ccccc}
\hline\hline   
JW & GOF & $\chi^2$ & Excess noise & d (pc) \\
\hline 
180 & 71.00 & 7652.0 & 2.22 & 323$\pm$35 \\
293 & 10.30 & 518.7 & 0.41 & 386$\pm$12 \\
647 & 0.32 & 164.0 & 0.04 & 412$\pm$7 \\
647 & 2.90 & 227.0 & 0.00 & 386$\pm$8 \\
908 & 17.80 & 807.0 & 0.56 & 395$\pm$14 \\
\hline 
\end{tabular}
\end{table}

In order to properly address the impact of assuming different distance values for JW180, we recalculated the stellar parameters for this target assuming an average distance to the ONC. We considered the distance of $414 \pm 7$ pc derived by \cite{menten2007} and compared the results with the previous ones computed with $323 \pm 35$ pc, derived from the Gaia DR2 parallax \citep{gaia2016,gaia2018,luri2018}. The stellar parameters derived for both distances are shown in Table \ref{table:stellarparams_JW180}.

\begin{table*}[]
\caption{Stellar parameters derived for JW180 considering different distance values ($d$) to derive the stellar parameters. The stellar parameters derived include effective temperature (\Teff), stellar luminosity (\Lstar), stellar radius (\Rstar), stellar mass (\Mstar), and age.}            
\label{table:stellarparams_JW180}     
\centering                         
\begin{tabular}{c c c c c c }       
\hline\hline                 
d (pc) & \Teff (K) & \Lstar (\Lsun) & \Rstar (\Rsun) & \Mstar (\Msun) & Age (Myr) \\  
\hline                        
   $323 \pm 35^{(1)}$ & 3125 $\pm$ 100 & 0.11 $\pm$ 0.05 & 1.12 $\pm$ 0.24 & 0.17 $\pm$ 0.04 & 3.53 $\pm$ 2.15 \\
   $414 \pm 7^{(2)}$  & 3125 $\pm$ 100 & 0.18 $\pm$ 0.07 & 1.43 $\pm$ 0.30 & 0.20 $\pm$ 0.04 & 2.38 $\pm$ 1.11 \\
\hline                                  
\end{tabular}
\tablebib{(1)~\citet{gaia2016,gaia2018,luri2018}; (2)~\citet{menten2007}.}
\end{table*}

\begin{figure}
   \centering
   \includegraphics[width=\hsize]{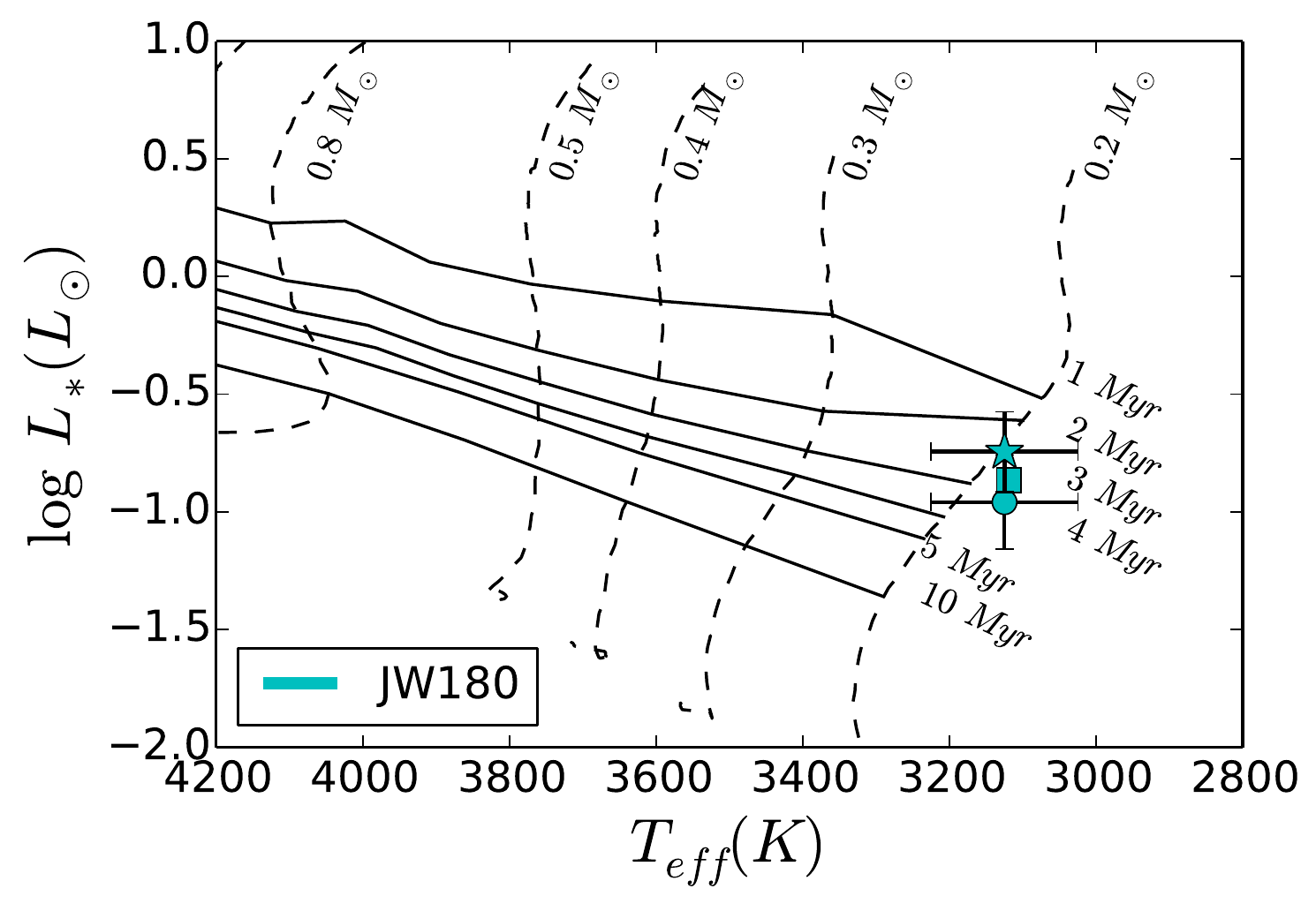}
      \caption{Hertzsprung-Russel diagram with \cite{siess2000} evolutionary tracks from 0.2 to 0.8 $M_\sun$ (dashed lines). Isochrones corresponding to 1, 2, 3, 4, 5, and 10 Myrs (solid lines) are also represented. The circle corresponds to the value determined in this work and the square to the value of \cite{hillenbrand1997}, rescaled to the distance derived from the Gaia DR2 parallax. The cyan star represents the value for JW180 assuming a distance of $414 \pm 7$ pc.}
         \label{HRD_Siess2000_JW180}
   \end{figure}

In Fig. \ref{HRD_Siess2000_JW180}, the cyan star represents the values recalculated for the averaged distance to the cluster. Compared to the previous value (cyan circle), JW180 appears to be 1.6 times brighter and $\sim1$ Myr younger. The small increment in both stellar radius and mass did not significantly affect the mean accretion rate for JW180. From the previous \Macc~of $10^{-10.1}M_\sun yr^{-1}$ we now get a value of $10^{-9.9} M_\sun yr^{-1}$ by considering 414 pc for the distance.

\section{Accretion tracers in emission}\label{app:accretion_tracers}

   \begin{figure*}[]
   \centering
   \includegraphics[width=\hsize]{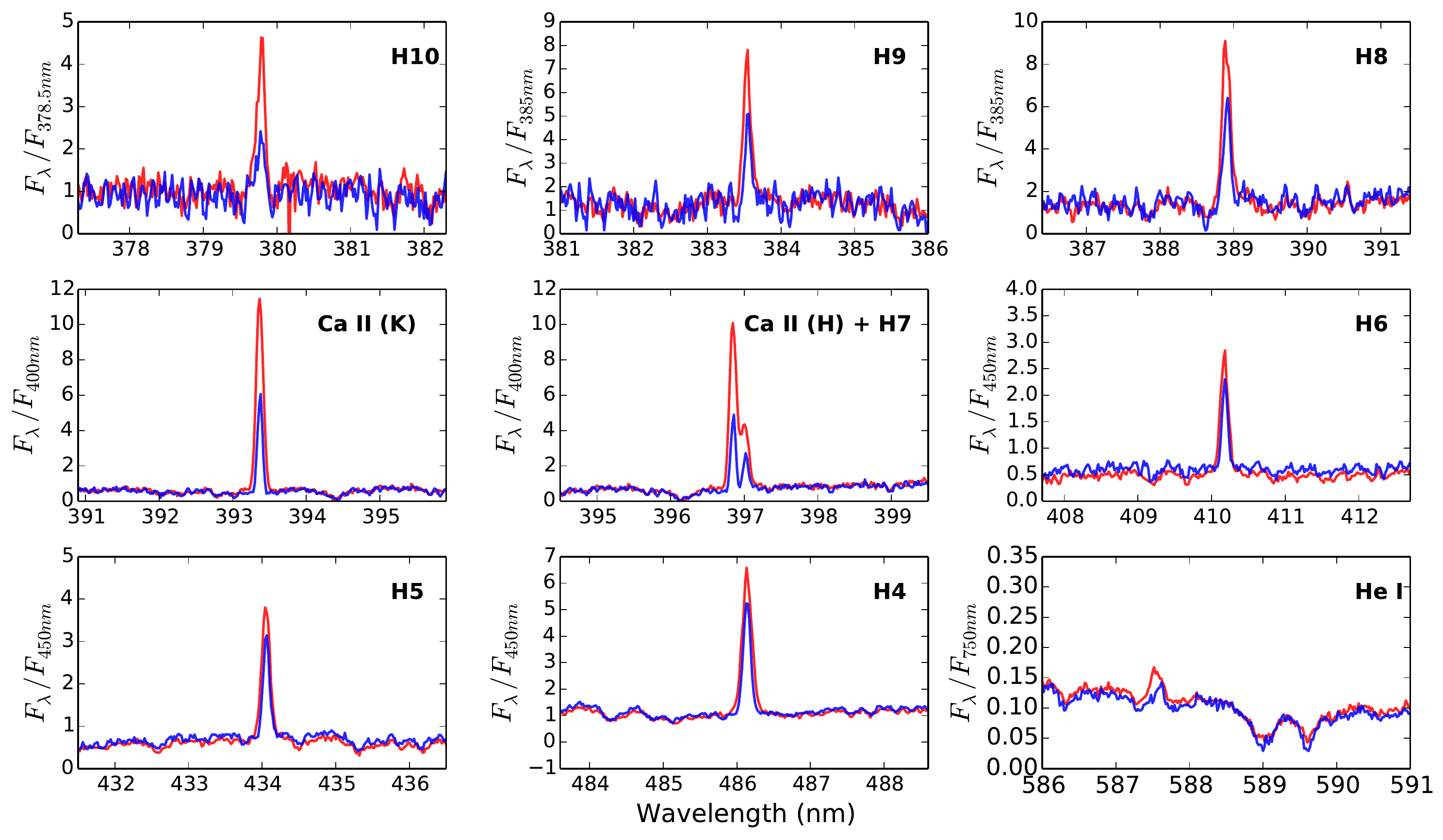}
      \caption{Accretion tracers in emission for JW180 (red line) and corresponding diskless template Par-Lup3-2 (blue line) taken from \cite{manara2013b}.}
         \label{fig:JW180}
   \end{figure*}

   \begin{figure*}[]
   \centering
   \includegraphics[width=\hsize]{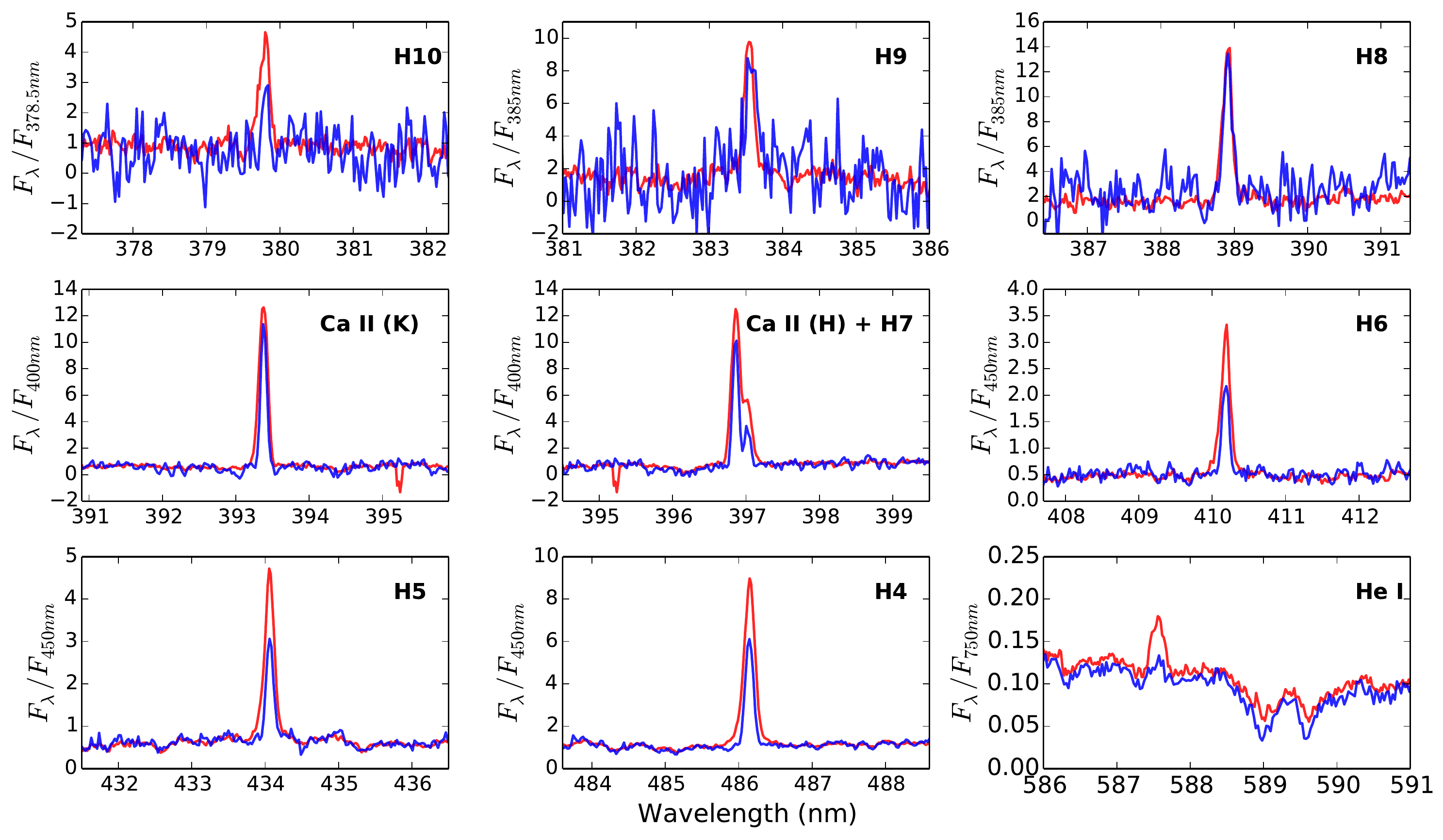}
      \caption{Accretion tracers in emission for JW293 (red line) and corresponding diskless template SO797 (blue line) taken from \cite{manara2013b}.}
         \label{fig:JW293}
   \end{figure*}

   \begin{figure*}[]
   \centering
   \includegraphics[width=\hsize]{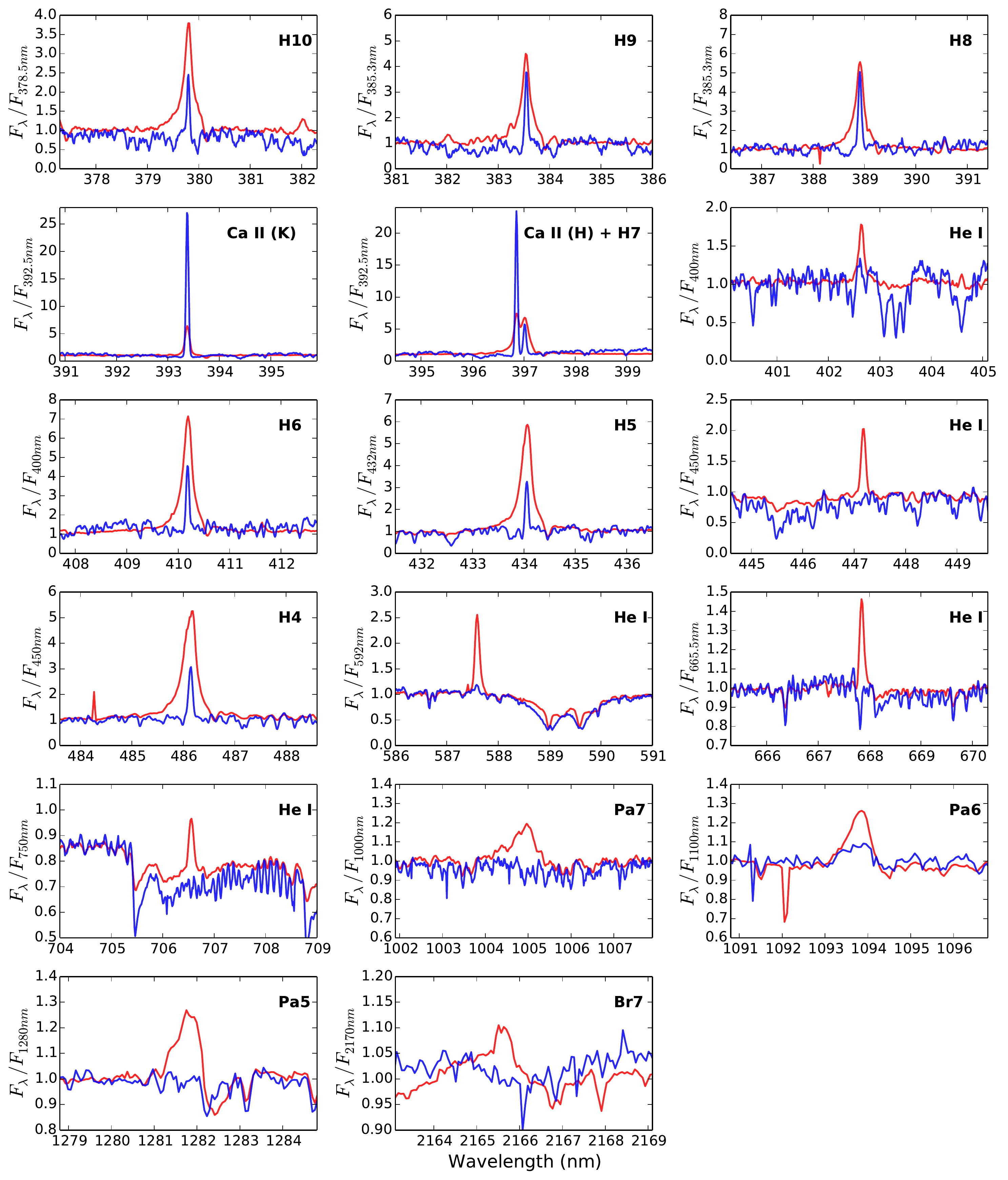}
      \caption{Accretion tracers in emission for JW647 (red line) and corresponding diskless template TWA13B (blue line) taken from \cite{manara2013b}.}
         \label{fig:JW647}
   \end{figure*}

   \begin{figure*}[]
   \centering
   \includegraphics[width=\hsize]{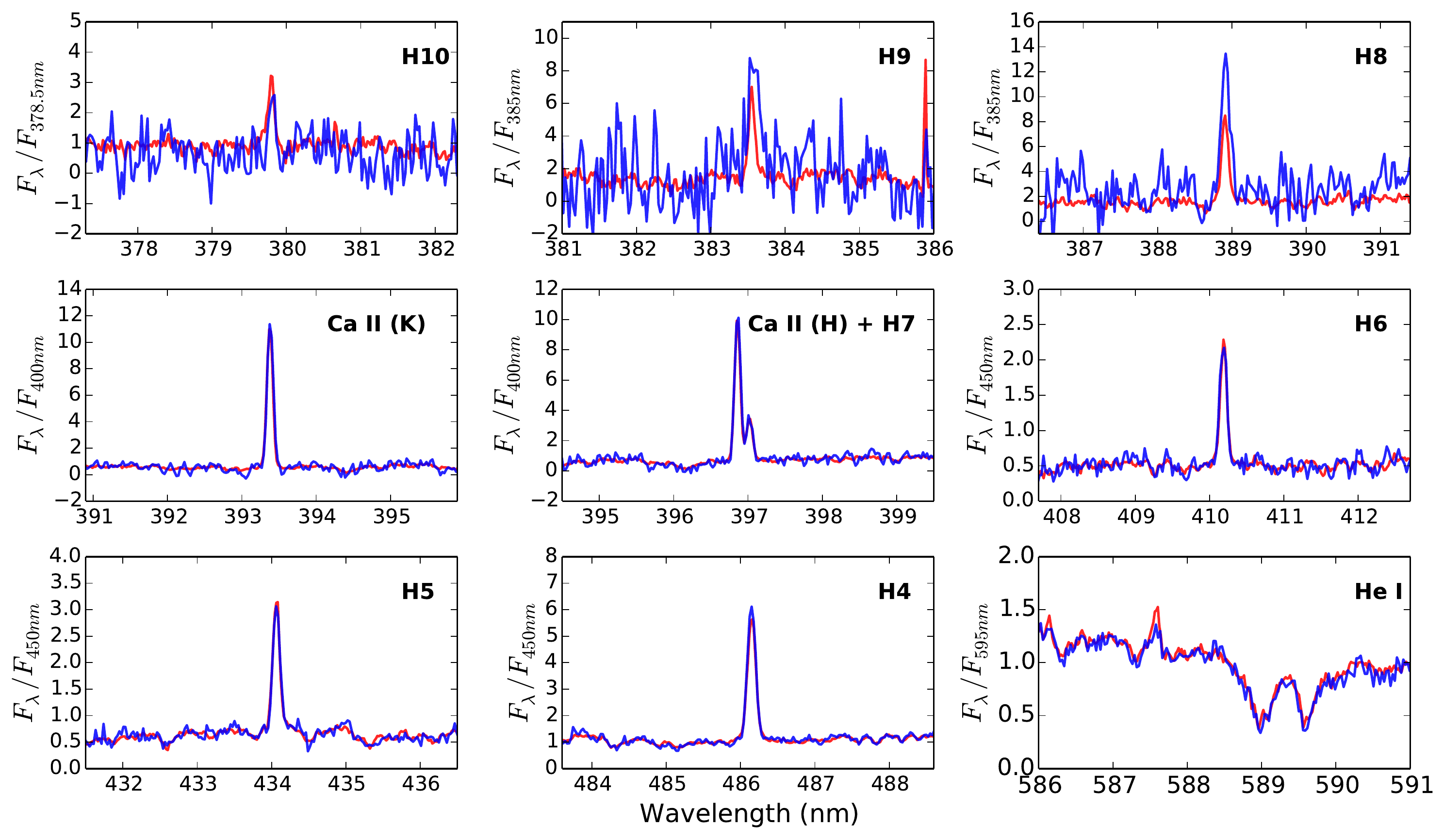}
      \caption{Accretion/chromospheric tracers in emission for JW908 (red line) and corresponding diskless template SO797 (blue line) taken from \cite{manara2013b}.}
         \label{fig:JW908}
   \end{figure*}

\end{appendix}

\end{document}